\documentclass[twocolumn,prb,superscriptaddress]{revtex4-1}
\usepackage[latin1]{inputenc}
\usepackage{graphicx}
\usepackage{grffile}
\usepackage{amssymb}
\usepackage{amsmath}
\usepackage{bbold}
\usepackage{xspace}
\usepackage{dcolumn}
\usepackage{bm}
\usepackage{color}
\usepackage{float}
\usepackage[extra]{tipa}

\setcitestyle{square,numbers}

\newfont{\tensy}{cmsy10}

\newcommand{\ie}[0]{i.e.\@\xspace}
\newcommand{\eg}[0]{e.g.\@\xspace}
\newcommand{\etal}[0]{\emph{et al.\@\xspace}}

\newcommand{\rmi}{\text{i}}

\newcommand{\oP}{\hat{P}}
\newcommand{\oQ}{\hat{Q}}

\newcommand{\on}{\hat{n}}

\newcommand{\rmd}{\text{d}}

\newcommand{\bit}{\begin{itemize}}
\newcommand{\eit}{\end{itemize}}

\newcommand{\kF}{k_\text{F}}
\newcommand{\kB}{k_\text{B}}
\newcommand{\nag}{{\phantom{\dag}}}

\newcommand{\bpsib}{\overline{\boldsymbol{\psi}}}
\newcommand{\bpsi}{{\boldsymbol{\psi}}}

\newcommand{\Kr}{K_\rho}

\newcommand{\las}[0]{\langle}
\newcommand{\ras}[0]{\rangle}

\newcommand{\Tr}[0]{\mbox{Tr}}

\newcommand{\im}{\mathrm{i}}

\renewcommand{\S}[0]{\mathcal{S}}
\newcommand{\SL}[0]{\mathcal{S}_{\text{L}}}
\newcommand{\SR}[0]{\mathcal{S}_{\text{R}}}
\newcommand{\SLL}[0]{\mathcal{S}_{\text{LL}}}
\newcommand{\SRR}[0]{\mathcal{S}_{\text{RR}}}

\newcommand{\nL}[0]{n_{\text{L}}}
\newcommand{\nR}[0]{n_{\text{R}}}
\newcommand{\nLL}[0]{n_{\text{LL}}}
\newcommand{\nRR}[0]{n_{\text{RR}}}

\newcommand{\Pp}{P_{\! +}}

\newcommand{\fcohan}[1]{c_{#1}}
\newcommand{\fcohcr}[1]{\bar{c}_{#1}}
\newcommand{\fcohmeasure}{\mathcal{D}(\bar{c},c)}

\newcommand{\expv}[1]{\left\langle #1 \right\rangle}
\newcommand{\expvtext}[1]{\langle #1 \rangle}

\begin{document}

%%%%%%%%%%%%%%%%%%%%%%%%%%%%%%%%%%%%%%%%%%%%%%%%%%%%%%%%%%%%%%%%%%%%%%%%%%%%%
%%%%%%%%%%%%%%%%%%%%% TITLE & ABSTRACT %%%%%%%%%%%%%%%%%%%%%%%
%%%%%%%%%%%%%%%%%%%%%%%%%%%%%%%%%%%%%%%%%%%%%%%%%%%%%%%%%%%%%%%%%%%%%%%%%%%%%

\title{Competing orders and unconventional criticality in the
  Su-Schrieffer-Heeger model}

\author{Manuel Weber}

\email[Please send correspondence to]{\ mw1162@georgetown.edu.}
\affiliation{\mbox{Department of Physics, Georgetown University, Washington,
    DC 20057, USA}}

\author{Francesco Parisen Toldin}

\affiliation{\mbox{Institut f\"ur Theoretische Physik und Astrophysik,
    Universit\"at W\"urzburg, 97074 W\"urzburg, Germany}}

\author{Martin Hohenadler}

\affiliation{\mbox{Institut f\"ur Theoretische Physik und Astrophysik,
    Universit\"at W\"urzburg, 97074 W\"urzburg, Germany}}

\begin{abstract}
  The phase diagram of the one-dimensional Su-Schrieffer-Heeger model of
  spinless fermions coupled to quantum phonons is determined by quantum Monte
  Carlo simulations. It differs significantly from previous work.
  In addition to Luttinger liquid and bond-order-wave (BOW) phases, we find
  an extended charge-density-wave (CDW) phase. Because of different broken symmetries, 
  BOW and CDW phases are connected by a retardation-driven phase transition.
  Our results are consistent with the theory of the frustrated $XXZ$ chain,
  including unconventional power-law exponents at criticality, and an
  interpretation in terms of deconfined quantum criticality via proliferation
  of solitons.
\end{abstract}

\date{\today}

\maketitle

\section{Introduction}

The rich phase diagrams of, \eg,  dichalcogenides \cite{manzeli20172d}
or cuprates \cite{RevModPhys.78.17}
motivate fundamental investigations of competing orders in strongly
correlated quantum systems. Motivated by the significant complexity of real
materials and justified by the concept of universality, most theoretical work
is based on minimal and hence tractable models. A recent focus are Dirac
systems \cite{sato2017dirac,PhysRevB.92.085147,li2017fermion,PhysRevB.95.085143,PhysRevB.97.041117,PhysRevB.97.081110,liu2019superconductivity}.
The existence of two or more ordered phases also provides a route to study non-Landau
deconfined quantum critical points (DQCPs), for which topological excitations
of the order parameters play a central role \cite{senthil2004decon}. For
recent progress, see Refs.~\cite{Nahum15,Shao15,wang2017decon,qin2017dual,liu2019superconductivity,PhysRevB.99.075103,mudry2019quantum,1DdcqpRoberts,1DdqcpHuang}.
Different orders can arise either from different interactions, or from local
and nonlocal components of the same Coulomb interaction \cite{Raghu08}. The
intricacy of such problems is reflected, \eg, in the debates surrounding the
complex phase diagrams of extended Hubbard models on one-dimensional (1D)
chains and 2D honeycomb lattices, see Refs.~\cite{MHHF2017,capponi2016phase}
for reviews. 

\begin{figure}[b]
  \includegraphics[width=0.425\textwidth]{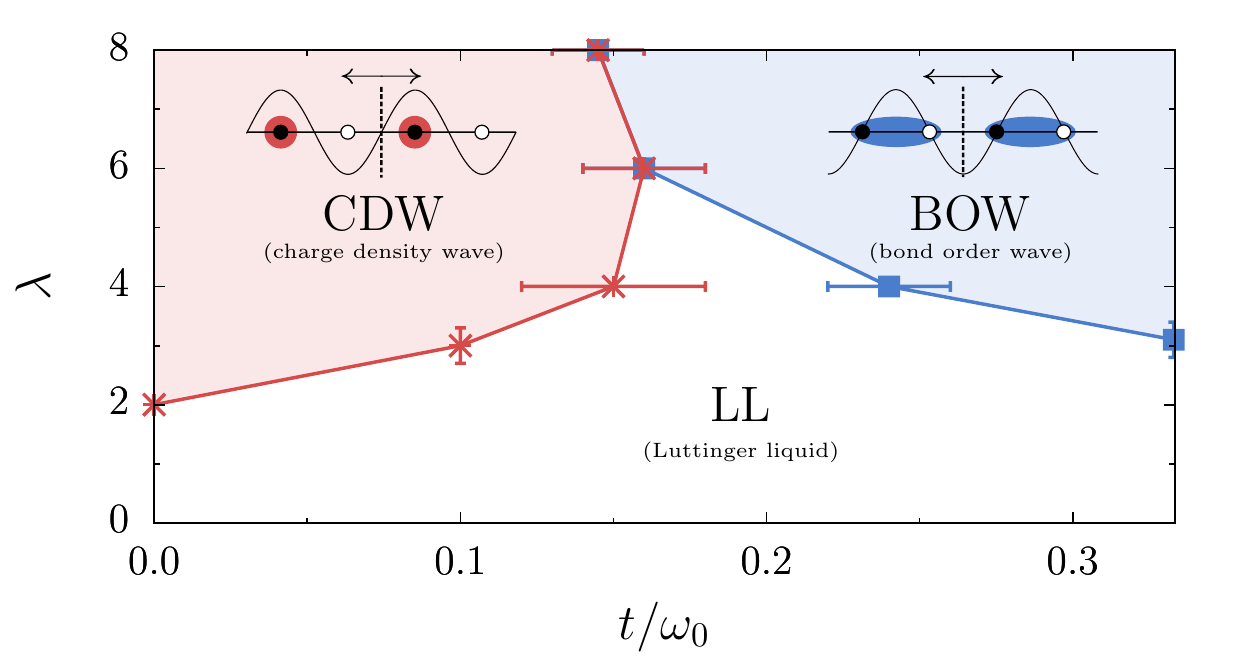}
  \caption{\label{fig:phasediagram} Phase diagram of the SSH
    model~(\ref{eq:SSH0}) as a function of inverse phonon frequency and
    electron-phonon coupling from QMC simulations.  Insets illustrate CDW/BOW
    order, $A/B$ sublattices, and inversion around the central bond.}
\end{figure}

Retardation effects, which are negligible for Coulomb interactions, play a
fundamental role in the context of electron-phonon coupling. As is known from
the theory of superconductivity \cite{PhysRev.108.1175}, phonon-mediated
interactions have attractive and repulsive components. However, for a
commensurate band filling, the different electron-phonon couplings are
commonly associated with only a single type of order each. A charge-density-wave (CDW) state with a
modulated electron site density follows from a Holstein coupling and is
observed in molecular crystals \cite{Pouget2016332}. A bond-order-wave (BOW) state with a
dimerized kinetic energy emerges from a Su-Schrieffer-Heeger (SSH) coupling
and is realized in conjugated polymers \cite{Pouget2016332}. Both orders are
illustrated in Fig.~\ref{fig:phasediagram} for a 1D chain.

The SSH
model \cite{PhysRevLett.42.1698} describes
electrons coupled to quantum phonons. 
Originally introduced to study topological solitons in
conjugated polymers \cite{PhysRevLett.42.1698}, SSH models also have close
relations with field theories of Dirac fermions \cite{fradkin2013field}. The
mean-field SSH model provides an important platform to explore interaction
and nonequilibrium effects on 1D topological phases
\cite{PhysRevB.96.165124,PhysRevLett.112.196404,PhysRevLett.121.090401,PhysRevB.98.214306,PhysRevLett.105.190403,bermudez2018gross,kuno2018generalized}. Novel
experimental realizations include cold atoms \cite{atala2013direct} and resistor networks \cite{lee2018topolectrical}. 

Despite the fundamental role of the SSH model, the
retarded nature and momentum dependence of the phonon-mediated
electron-electron interaction have so far prohibited a detailed understanding.
Compared to simpler models of spins or fermions, theoretical and numerical
approaches to electron-phonon problems face significant challenges
\cite{MHHF2017}. Motivated by this situation, we explore the full phase
diagram of the 1D spinless SSH model by quantum Monte Carlo (QMC) simulations and find
important deviations from existing work and rich, previously undiscovered physics.
We demonstrate the existence of not one
\cite{PhysRevB.27.1680} but two {\em dimerized} phases upon variation of the
phonon frequency $\omega_0$: the familiar BOW
Peierls phase with dimerized
hopping and an extended
CDW phase.
This raises the central question of 
the nature of the BOW-CDW phase transition, which is entirely beyond the
usual
adiabatic and antiadiabatic approximations.
Our findings, most notably
unconventional exponents at criticality, are consistent with a field theory
previously developed for frustrated spin chains \cite{PhysRevB.25.4925}. They
also establish 1D DQCP physics \cite{mudry2019quantum} in an electron-phonon model.

The paper is organized as follows. In Sec.~\ref{Sec:Model} we define the SSH model
and discuss its exact limits, in Sec.~\ref{Sec:Results} we present our QMC results,
in Sec.~\ref{Sec:Discussion} we discuss our results, and in Sec.~\ref{Sec:Conclusions}
we conclude. The Appendices contain details on the QMC method and additional data.

\section{Model \& exact limits\label{Sec:Model}}

The Hamiltonian of the SSH model
\cite{PhysRevLett.42.1698,PhysRevB.67.245103},
\begin{align}\label{eq:SSH0}
  \hat{H}
  &=
    -t \sum_{b} \hat{B}_{b} 
    +
    g
    \sum_{b} 
    \hat{B}_{b}
    \oQ_{b}
    +
    \hat{H}_\text{ph}
    \,,
\end{align}
with
$\hat{B}_{b} = \hat{c}^\dag_{i(b)} \hat{c}^\nag_{j(b)} + \hat{c}^\dag_{j(b)}
\hat{c}^\nag_{i(b)}$
acting on bond $b$ between sites $i$ and $j=i+1$ and
$\hat{H}_\text{ph}=\sum_b \left(\frac{1}{2M}\oP^2_{b} + \frac{K}{2}
  \oQ_{b}^2\right)$,
describes spinless fermions coupled to
optical bond phonons with momentum $\oP_b$, displacement $\oQ_b$, and
frequency $\omega_0=\sqrt{K/M}$. For the present case of a half-filled band
($\las\on^{}_i\ras=\las \hat{c}^\dag_i \hat{c}^\nag_i\ras=0.5$), numerics
\cite{PhysRevB.91.245147,Bakrim2015} and field theory \cite{PhysRevB.27.1680}
suggest the same physics for optical phonons---which preclude a QMC sign
problem---and the original acoustic phonons \cite{PhysRevLett.42.1698}. 
After integrating out the phonons, the partition function 
contains the retarded interaction
\begin{equation}\label{eq:S1} {S}_\text{ret} = -\frac{\lambda
    t}{2}\iint_0^\beta \rmd \tau \rmd \tau' \sum_{b} B_b(\tau)
  P(\tau-\tau') B_b(\tau')\,.
\end{equation}
The free phonon propagator $P(\tau)$ is local in space but its decay in
imaginary time $\tau$ (here, $\beta=1/T$) is determined by $\omega_0$, with
$P(\tau)\sim e^{-\omega_0\tau}$. The associated retardation effects are
crucial for the phase diagram in Fig.~\ref{fig:phasediagram}. We
use the coupling $\lambda=g^2/Kt$ and set $\hbar=\kB=1$.

For $\omega_0=0$, corresponding to classical phonons, mean-field theory is
exact at $T=0$. Replacing $\oQ_b$ with $\las\oQ_b\ras = (-1)^b\Delta/g$ in
Eq.~(\ref{eq:SSH0}) yields the fermionic hopping term
$\hat{H}_0 = -\sum_b [t + (-1)^b\Delta] \hat{B}_b$. The Peierls argument \cite{Peierls} implies that the
bond dimerization $\Delta$ is nonzero for any $\lambda>0$ and opens a gap at
the Fermi level. Quantum lattice fluctuations can destroy long-range order at
sufficiently weak coupling and thereby allow for a Luttinger liquid (LL) to
BOW quantum phase transition (QPT) at a finite $\lambda_c(\omega_0)$
\cite{PhysRevB.27.1680,Ba.Bo.07,PhysRevB.91.245147}. An exact solution (by
the Bethe ansatz) is also possible in the opposite, antiadiabatic limit.  For
$\omega_0\to\infty$, the interaction~(\ref{eq:S1}) becomes instantaneous and
Eq.~(\ref{eq:SSH0}) maps to the $t$-$V$ model $\hat{H}_\infty=-t\sum_b
\hat{B}_b + V \sum_i \on_i \on_{i+1}$ with $V=\lambda$
\cite{PhysRevB.27.1680} and a LL-CDW QPT at $V_c/t=\lambda_c=2$
\cite{shankar1990solvable}. 

BOW and CDW states, illustrated in Fig.~\ref{fig:phasediagram}, {\em spontaneously} break translation symmetry and are
described by Ising order parameters that reflect the two possible BOW (CDW)
dimerization patterns related by a shift by one lattice constant. The discrete
Ising symmetry permits long-range order at $T=0$. CDW and BOW states can be distinguished by point
group symmetries. CDW order
breaks bond inversion symmetry but preserves invariance under site
inversion, see Fig.~\ref{fig:phasediagram}. The opposite is true for the BOW
phase. Given the different broken symmetries, a phase transition
is expected between $\omega_0=0$ and $\omega_0=\infty$. In contrast, the
influential Ref.~\cite{PhysRevB.27.1680} suggests a single dimerized phase
mainly based on a continuum theory, details and extensions of which 
are discussed below. Similar conclusions were reached by (functional) RG
calculations \cite{PhysRevB.29.4230,Ba.Bo.07} and for related spin-phonon
models \cite{PhysRevLett.76.4050,PhysRevB.56.14414,Ci.Or.Gi.05,Ba.Bo.07}. A
nonadiabatic mean-field approach yields BOW {\em and} CDW phases even for
large $\omega_0$ \cite{sil1998spin}. These findings differ significantly from
our results. Finally, the choice of spinless fermions is motivated by the
absence of a LL phase \cite{MHHF2017} and only one symmetry-broken (BOW)
phase in the spinful case; at small $t/\omega_0$, critical spin correlations
are expected.

\section{Results\label{Sec:Results}}

Our simulations were made possible by using a state-of-the-art
QMC method based on the stochastic series expansion \cite{PhysRevB.43.5950}
and directed-loop updates \cite{Sandvik02}.
The performance gain from extending the latter to retarded interactions \cite{arXiv:1704.07913}
is essential to explore the phase diagram of the SSH model. The method
has only statistical errors and relevant technical details are summarized
in App.~\ref{App:QMC}.
Results were obtained for
periodic chains of $L$ sites, and for inverse temperatures $\beta t=2L$
representative of $T=0$.

\subsection{Identification of the different phases}

\begin{figure}
  \includegraphics[width=0.45\textwidth]{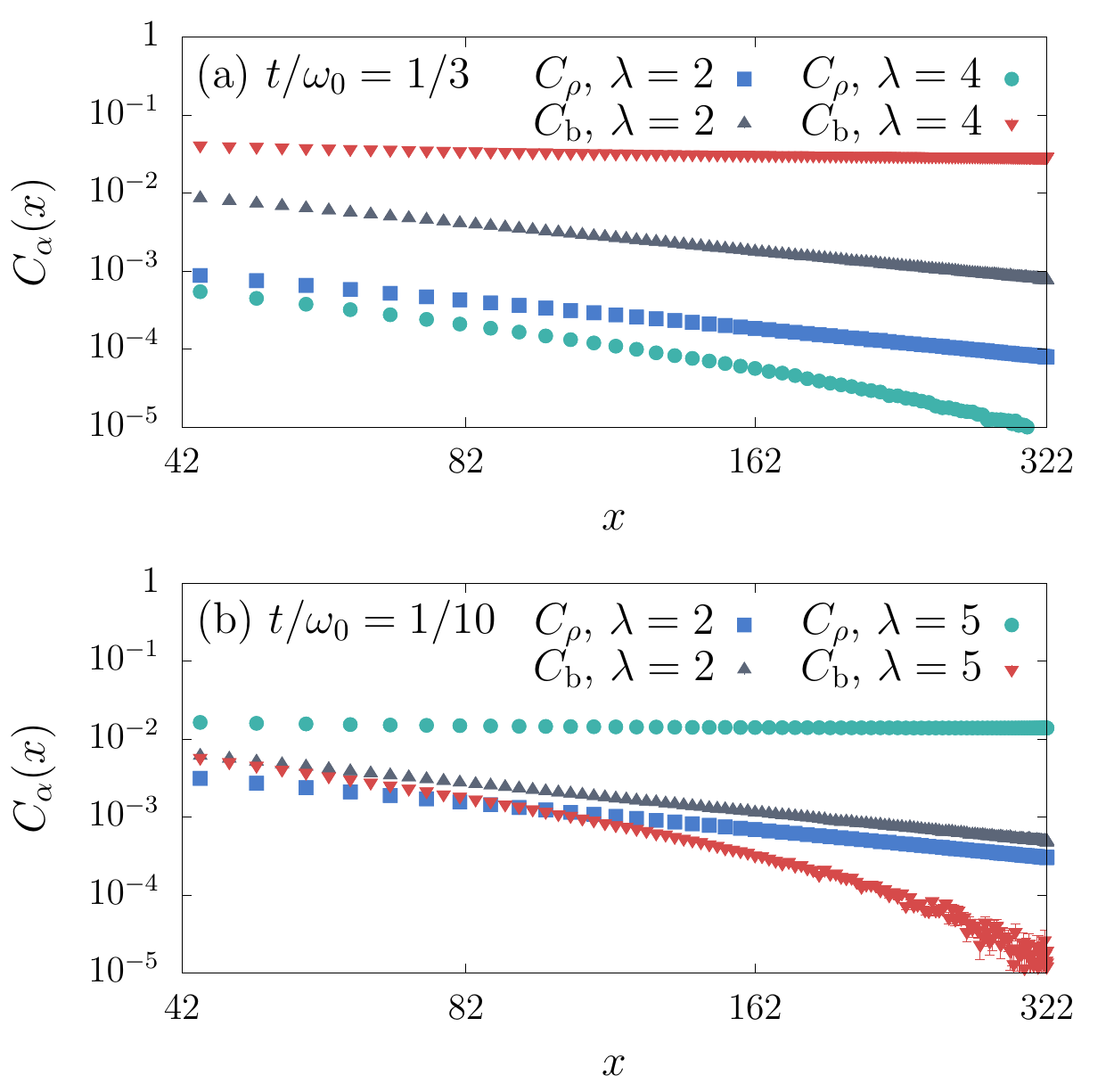}
  \caption{\label{fig:correlators} Real-space charge ($\alpha=\rho$) and bond
    ($\alpha=\text{b}$) correlation functions. Here, $L=322$, $\beta t=2L$,
    and $x=L\sin \left(\pi r/L\right)$.}
\end{figure}

The three distinct regimes in Fig.~\ref{fig:phasediagram}
are revealed by the real-space correlation functions. Figure~\ref{fig:correlators}
shows the charge ($\rho$) and bond ($\text{b}$) correlators
\begin{align}\nonumber
  C_{\rho}(r) &=  \las (\on_r -\las \on_r\ras) (\on_0-\las n_0\ras) \ras\,,\\\label{eq:correlators}
  C_\text{b}(r) &= \las (\hat{B}_{r}-\las \hat{B}_r\ras)(\hat{B}_{0}-\las \hat{B}_0\ras)\ras\,.
\end{align}
Only positive values are visible in the logarithmic
representation.
Using the conformal distance $x=L\sin \left(\pi r/L\right)$ minimizes the
effects of the periodic boundaries. For
$t/\omega_0=1/3$ and $\lambda=2$ [Fig.~\ref{fig:correlators}(a)], both
exhibit a power-law decay of $q=2\kF$ correlations described by the LL
expressions for 1D metals \cite{Giamarchi}
\begin{align}
\nonumber
C_\rho(r) &= -\frac{\Kr}{2\pi^2r^2} + \frac{A_\rho}{r^{2\Kr}}  \cos(2\kF r) \, , \\
C_\text{b}(r) &=\frac{A_\text{b}}{r^{2\Kr}} \cos(2\kF r) \, ,
\label{Eq:CorLL}
\end{align}
with $\Kr\approx 0.61$.
At $\lambda=4$, corresponding to the BOW phase in Fig.~\ref{fig:phasediagram},
the saturation of $C_\text{b}(x)$ at large $x$ and the exponential decay of
$C_{\rho}(x)$ are consistent with long-range bond order.
In the nonadiabatic regime [$t/\omega_0=1/10$,
Fig.~\ref{fig:correlators}(b)], we find behavior consistent with a LL at
$\lambda=2$ and long-range CDW order at $\lambda=5$.

\begin{figure}
  \includegraphics[width=0.45\textwidth]{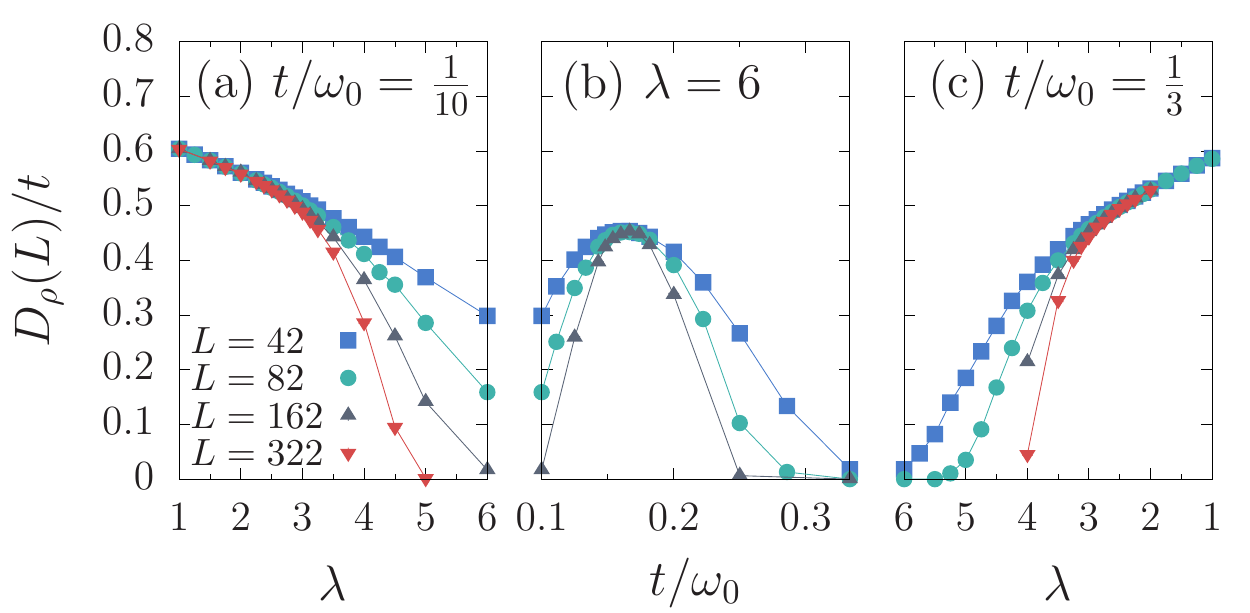}
  \caption{\label{fig:stiffness}
  Charge stiffness $D_\rho$ along a continuous path 
  through the phase diagram in Fig.~\ref{fig:phasediagram}
  which captures (a) the LL-CDW
  transition at $t/\omega_0 = 1/10$, (b) the CDW-BOW transition at $\lambda=6$, and 
  (c) the BOW-LL transition at $t/\omega_0 = 1/3$.
   $D_\rho(L)$ is nonzero for $L\to\infty$ in the
    metallic LL phase and at the CDW-BOW transition point.
    }
\end{figure}

A general diagnostic to distinguish metallic and insulating 1D
phases is the charge stiffness
\begin{align}
\label{Eq:Drude}
  D_\rho = L \, \frac{\partial^2 E(\phi)}{\partial\phi^2} \bigg\rvert_{\phi=0} \,,
\end{align}
which measures the response of the
system---via the ground-state energy $E$---to a magnetic
flux $\phi$ \cite{PhysRev.133.A171}.
In our simulations, it can be obtained from the estimator of the superfluid stiffness
at low enough temperatures, see App.~\ref{App:Stiffness}. 
For $L\to\infty$, the charge stiffness is nonzero (zero) in metallic (insulating) phases
\cite{Scalapino93,PhysRevB.65.155113}.
To illustrate the topology of the phase diagram, we show in
Fig.~\ref{fig:stiffness} the stiffness $D_\rho(L)$ along a continuous path in
parameter space:
(a) we first increase $\lambda$ from 1 to 6 at fixed $t/\omega_0=1/10$, (b) then we
tune $t/\omega_0$ from $1/10$ to $1/3$ at fixed $\lambda=6$, and (c) finally we
decrease $\lambda$ again from 6 to 1 at fixed $t/\omega_0=1/3$.
In accordance with Fig.~\ref{fig:phasediagram}, $D_\rho(L)$
remains nonzero for $\lambda< 3$ (LL phase) and clearly vanishes for
$L\to\infty$ for $\lambda\geq 4$ (CDW phase) at fixed $t/\omega_0=1/10$, see  
Fig.~\ref{fig:stiffness}(a). The determination of critical values and the
finite-size scaling at and near $\lambda_c$ ($\lambda_c\approx 3$ according to Fig.~\ref{fig:phasediagram})
will be discussed in detail below [see also Fig.~\ref{fig:stiffness-criticalvalues}(a)].
Similar behavior is observed in Fig.~\ref{fig:stiffness}(c) for the LL-BOW transition
at $t/\omega_0=1/3$.
Remarkably, at fixed $\lambda=6$, the stiffness converges to a
nonzero value only at intermediate $t/\omega_0$, see Fig.~\ref{fig:stiffness}(b).
It is shown in App.~\ref{App:AddDataOrder} that the
peak in $D_\rho(L)$ narrows with increasing $\lambda$ but its maximum value remains finite.
Naively, one would assume that the insulating BOW and CDW phases are separated
by an extended metallic region. However, numerical results
and field-theory arguments discussed below
provide evidence for a metallic line separating the ordered phases, as suggested by Fig.~\ref{fig:phasediagram}.

\subsection{Finite-size scaling of the charge stiffness}

\begin{figure}
  \includegraphics[width=0.45\textwidth]{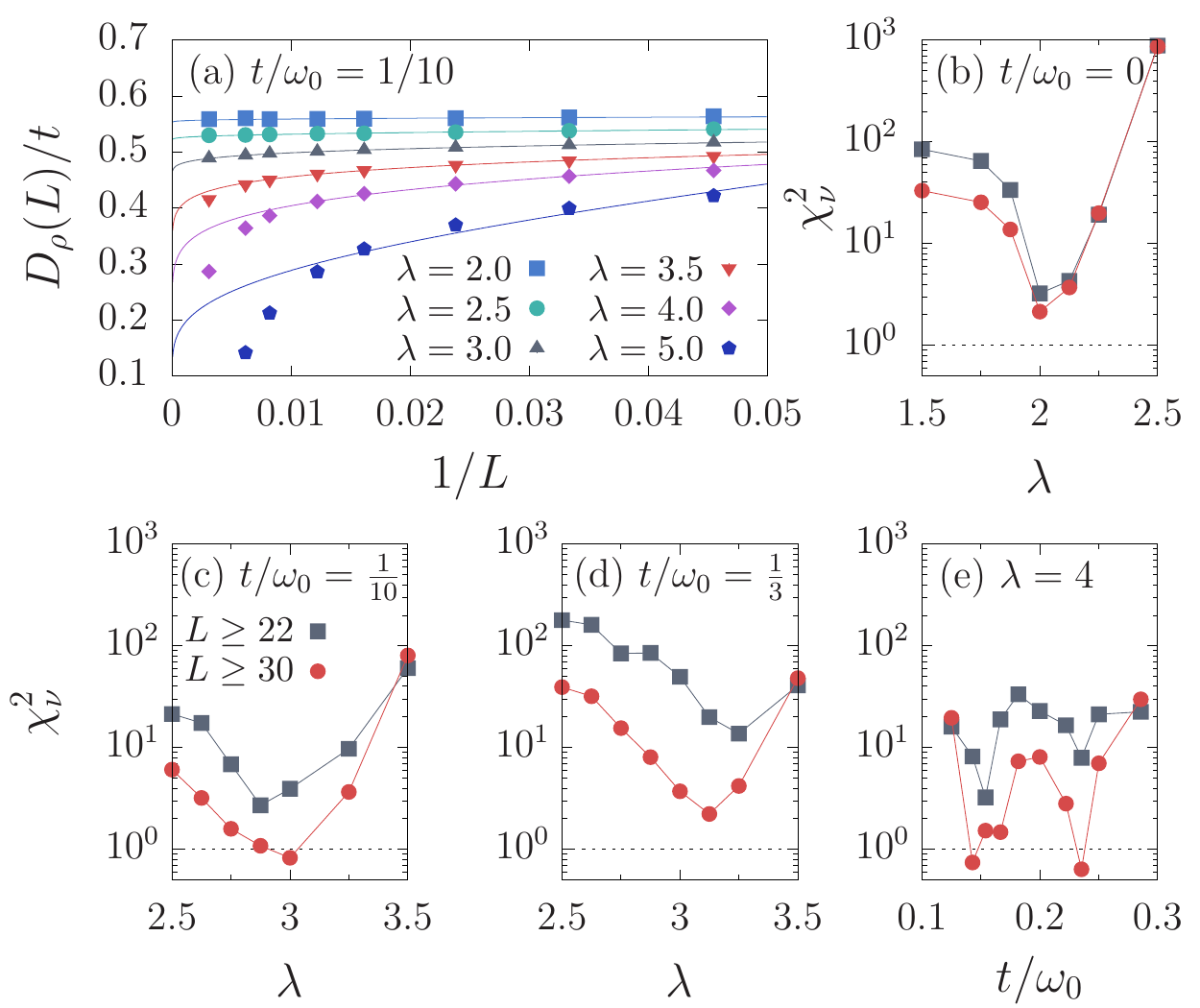}
  \caption{\label{fig:stiffness-criticalvalues} 
    (a) Finite-size scaling of the charge stiffness for the case of $t/\omega_0=1/10$.
    (b)--(e) Reduced $\chi^2$ for fits of $D_\rho(L)$ to
    Eq.~(\ref{eq:WeberMinnhagen}) for different values of $t/\omega_0$. As
    explained in the text, the minima provide estimates for critical values.
 }
\end{figure}

To obtain the LL-BOW and LL-CDW critical values in Fig.~\ref{fig:phasediagram}, we analyzed
the finite-size scaling of $D_\rho(L)$, see Fig.~\ref{fig:stiffness-criticalvalues}(a).
For 1D metal-insulator transitions, a renormalization-group (RG) analysis of
umklapp interactions predicts numerically challenging
Berezinskii-Kosterlitz-Thouless (BKT) scaling with a critical value
$K_\rho=1/2$ \cite{Giamarchi}. This scenario has been
explicitly confirmed for $t/\omega_0=0$ \cite{shankar1990solvable}. Interestingly,
a functional RG study of the SSH model \cite{Ba.Bo.07} reported unconventional
BKT physics with $K_\rho< 1/2$. 
Large-scale simulations of classical frustrated 2D $XY$ models
\cite{HPV-05c,Kosterlitz_2016}---relevant due to the usual quantum-classical
mapping and a connection between retardation and frustration explained
below---indicate a standard BKT transition \cite{PhysRevB.72.184502,HPV-05c}
(see, however, Ref.~\cite{lima2018fully}), albeit with challenging crossover phenomena.

\begin{figure}
  \includegraphics[width=0.45\textwidth]{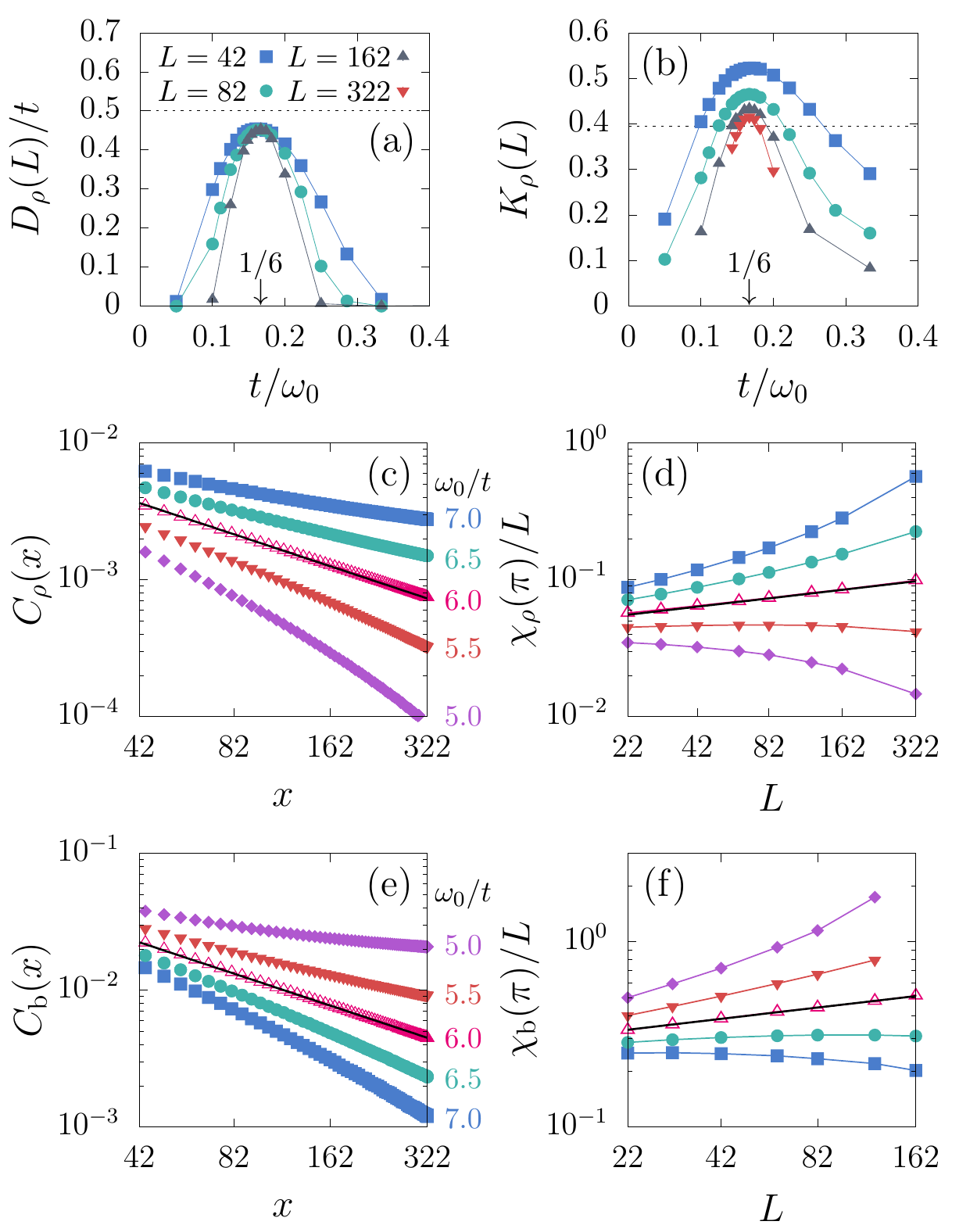}
  \caption{\label{fig:lambda6} Results for the BOW-CDW transition at $\lambda=6$.
    The saturation with system size
    of (a) the charge stiffness and (b) the LL parameter indicates metallic
    behavior near $t/\omega_0=1/6$.
    The latter is reflected in the power-law decay of real-space charge and
    bond correlations with the same exponent
    $\overline{\eta}=2K_\rho\approx0.79$ (solid black lines)
    in (c), (e) observed at fixed $L=322$. The exponent
    $\overline{\eta}$ also describes the
    scaling of the staggered charge/bond susceptibilities
    $\chi_{\rho/\mathrm{b}}(\pi)/L \sim L^{1-\overline{\eta}}$ in
    (d), (f). Here,  keys apply left and right. Data for
    $L=322$ are absent in (a) because converged results for $D_\rho$ are
    more challenging to obtain than for $K_\rho$.}
\end{figure}

The RG calculations predict a characteristic logarithmic scaling exactly
{\em at a critical point} due to marginally relevant operators. To first order \cite{PhysRevB.37.5986}, 
\begin{equation}\label{eq:WeberMinnhagen}
  \frac{D_\rho(L)}{D_\rho(\infty)}
  = 
  1 + \frac{g}{2\ln L + C}\,.
\end{equation}
As demonstrated before for 2D classical $XY$ models
\cite{PhysRevB.37.5986,hasenbusch2005two} and 1D quantum
models \cite{PhysRevLett.113.260403,gerster2016superfluid},
critical values can be extracted by fitting the stiffness data to
Eq.~(\ref{eq:WeberMinnhagen}), treating the jump $D_\rho(\infty)$
\cite{PhysRevLett.39.1201} as well as $g$ and $C$ as parameters.
Because the form~(\ref{eq:WeberMinnhagen})
only holds at critical points, the latter can be identified as the minima
in standard goodness-of-fit measures, such as the reduced chi-squared
$\chi_\nu^2=\chi^2/\nu$ for $\nu$ degrees of freedom used here.
Technical details of the stiffness fits are discussed in App.~\ref{App:StiffnessFits}.
The universal stiffness jump at the critical point of the 2D $XY$ model
\cite{PhysRevLett.39.1201} translates to $D_\rho(\infty)=t/2$ for the $t$-$V$
model (the SSH model with $t/\omega_0=0$) \cite{shankar1990solvable}. For
$t/\omega_0>0$, we instead find nonuniversal stiffness jumps, with
$D_\rho(L)<t/2$ even for small $L$ in, \eg,
Fig.~\ref{fig:lambda6}(a). Possible origins 
are discussed in App.~\ref{App:StiffnessJumps}.

For $t/\omega_0=0$ (the $t$-$V$ model), such fits indeed yield a minimum of
$\chi^2_\nu$ at the exact critical value $\lambda_\text{c}=2$ in
Fig.~\ref{fig:stiffness-criticalvalues}(b). For $t/\omega_0=1/10$, we
estimate $\lambda_\text{c}=3.0(3)$ from
Fig.~\ref{fig:stiffness-criticalvalues}(c).  Finite-size effects are larger
for a smaller $\omega_0$ (longer interaction range in $\tau$), as visible in
Fig.~\ref{fig:stiffness-criticalvalues}(d) for $t/\omega_0=1/3$ in terms of a
larger minimal $\chi^2_\nu$; the critical value is determined as
$\lambda_\text{c}=3.1(3)$. At fixed $\lambda=4$
[Fig.~\ref{fig:stiffness-criticalvalues}(d)], the stiffness fits exhibit two
separate minima at $t/\omega_{0,\text{c1}}=0.15(3)$ and
$t/\omega_{0,\text{c2}}=0.24(2)$, respectively, consistent with two critical
points.  For $\lambda=6$ and $8$, due to restrictions in system size, we
estimated a single critical value from the peaks in $D_\rho(L)$ and
$K_\rho(L)$ (see Fig.~\ref{fig:lambda6} and App.~\ref{App:AddData}) as
$t/\omega_{0,\text{c}}=0.16(2)$ and $t/\omega_{0,\text{c}}=0.145(15)$,
respectively.
These critical values appear consistent with the
finite-size behavior of other observables, as discussed
in App.~\ref{App:AddData}.

\subsection{CDW-BOW transition\label{Sec:CDW-BOW}}

Results for the retardation-driven CDW-BOW transition, a
key feature of the SSH model, are presented in Fig.~\ref{fig:lambda6} for
$\lambda=6$.  The convergence of $D_\rho(L)$ in Fig.~\ref{fig:lambda6}(a) and
$K_\rho(L)=2\pi C_\rho(q_1)/q_1$ (with $q_1=2\pi/L$) in
Fig.~\ref{fig:lambda6}(b) to a nonzero value around
$t/\omega_{0,\text{c}}\approx 1/6$ indicates a metallic state. However, $K_\rho(L)$ falls below the typical lower bound $1/2$ for
a LL \cite{Giamarchi}. The dotted line in Fig.~\ref{fig:lambda6}(b) is
$K_\rho=\overline{\eta}/2$, with $\overline{\eta}=0.79$ the exponent of the real-space charge
and bond correlations at $t/\omega_{0}=1/6$ in
Figs.~\ref{fig:lambda6}(c) and \ref{fig:lambda6}(e). Combining the LL form of
$C_\alpha(r)$, $\alpha \in \{\rho,\mathrm{b}\}$, given by
Eq.~(\ref{Eq:CorLL})
with conformal invariance implies
$\chi_\alpha(\pi)/L \sim L^{1-\overline{\eta}}$ \cite{PhysRevB.92.245132}
for the staggered charge/bond susceptibility 
\begin{align}
\label{Eq:Sus}
\chi_\alpha(\pi) = \sum_{r} (-1)^{r} \int_0^\beta d\tau \, C_\alpha(r,\tau) \, .
\end{align}
That this relation holds at $t/\omega_0=1/6$ [Figs.~\ref{fig:lambda6}(d) and \ref{fig:lambda6}(f)] is evidence
for a gapless state described by a conformal field theory. In particular,
charge and bond susceptibilities have the same power-law exponent (within the
available accuracy) and do not show any
visible finite-size corrections even down to the
smallest system sizes considered.
Away from
$t/\omega_0=1/6$, Figs.~\ref{fig:lambda6}(c)--\ref{fig:lambda6}(f) suggest either BOW or CDW order
with one of the orders instantly becoming dominant and the other being suppressed.
Our findings provide evidence for a direct quantum phase transition between CDW and BOW order with
an intermediate metallic point instead of a narrow region. This
interpretation is further supported
by theoretical arguments given in Sec.~\ref{Sec:Discussion}.
Data on the CDW-BOW transition for other $\lambda$
can be found in App~\ref{App:AddData}.

\section{Discussion\label{Sec:Discussion}}

The retardation effects that drive the
BOW-CDW transition are generally difficult to capture analytically
\cite{Voit86,Ba.Bo.07,Bakrim2015}. However, we find our numerical results to
be fully consistent with a field theory previously derived for the antiferromagnetic
$J_1$--$J_2$ $XXZ$ chain with Hamiltonian
$
  \hat{H}_{J_1-J_2} = J_1 \hat{H}_{XXZ}  
  + J_2 \sum_i {\boldsymbol{S}}_i {\boldsymbol{S}}_{i+2}
 $
and $\hat{H}_{XXZ}=\sum_i(\hat{S}^x_i\hat{S}^x_{i+1} + \hat{S}^y_i\hat{S}^y_{i+1} + {\Delta_z}\hat{S}^z_i\hat{S}^z_{i+1})$
\cite{PhysRevB.25.4925}.
Its bosonized continuum description takes the usual Tomonaga-Luttinger form \cite{PhysRevB.25.4925,mudry2019quantum}
\begin{equation}\label{eq:bosonized}
  \mathcal{H} = \frac{v}{2}\left[
    \frac{1}{\eta} \,(\partial_x \theta)^2 + {\eta}(\partial_x \phi)^2 + \lambda_\phi\cos(\sqrt{8\pi}\phi)
  \right]\,.
\end{equation}
Here, $v$ is the renormalized velocity, $\eta=1/\overline{\eta}=1/(2K_\rho)$ determines the exponent of
correlation functions, and
the cosine term encodes umklapp scattering. The phase diagram of
$\hat{H}_{J_1-J_2}$
has the same topology as Fig.~\ref{fig:phasediagram}
\cite{PhysRevB.25.4925}; the N\'{e}el and dimer phases correspond to
CDW and BOW phases, respectively.

A connection between the frustrated $XXZ$ model $\hat{H}_{J_1-J_2}$
and the SSH
model~(\ref{eq:SSH0}) can be established via a mapping of phonon-mediated
retarded interactions at small $t/\omega_0$ to frustrated spin interactions
\cite{kuboki1987spin,PhysRevB.65.144438,Ci.Or.Gi.05,PhysRevB.74.214426}, but
is expected to contain additional terms not present in 
$\hat{H}_{J_1-J_2}$. Alternatively, a Hamiltonian of the form~(\ref{eq:bosonized}) can in
principle be obtained for the SSH model via an RG treatment of the phonon-mediated interaction.
In either case, in contrast to the $XXZ$ problem \cite{PhysRevB.25.4925,mudry2019quantum},
an explicit expression for $\lambda_\phi$ in terms of the SSH model
parameters is not known. Here, we focus on comparing the predictions for
Eq.~(\ref{eq:bosonized}) \cite{PhysRevB.25.4925,mudry2019quantum} to our numerical data.

The LL-BOW and LL-CDW transitions can be attributed to the cosine umklapp
term in Eq.~(\ref{eq:bosonized}) becoming relevant for $K_\rho<1/2$.
In contrast to approximate functional RG results \cite{Ba.Bo.07}, we do not find evidence for
a violation of the conventional $K_\rho = 1/2$ at the critical point (see App.~\ref{App:AddDataLL} for details).
BOW and CDW phases are associated with opposite signs of $\lambda_\phi$ and a
pinning of the charge mode $\phi$ at different minima \cite{Giamarchi,mudry2019quantum}. 
The CDW phase is known in the Gross-Neveu
literature as an Aoki phase \cite{PhysRevD.30.2653}.
The line of BOW-CDW transitions suggested by Fig.~\ref{fig:phasediagram} for
$\lambda\gtrsim 6$ mirrors the line of continuous dimer-Néel transitions of
the frustrated $XXZ$ chain, along which $K_\rho$ varies continuously \cite{PhysRevB.25.4925,mudry2019quantum}.
Within the theory~(\ref{eq:bosonized}), and given $K_\rho<1/2$ [see Fig.~\ref{fig:lambda6}(b)], a metallic state
separating BOW and CDW phases requires $\lambda_\phi=0$. Since $\lambda_\phi$ depends on the independent
SSH model parameters $t/\omega_0$ and $\lambda$, it will vanish at a single value $(t/\omega_0)_\text{c}$ for a given
$\lambda$. 
This scenario is fully supported by our observation that for a given
$\lambda\gtrsim 6$, bond and charge correlations
show exactly the same power-law exponents at a single $(t/\omega_0)_\text{c}$
in Figs.~\ref{fig:lambda6}(c)--\ref{fig:lambda6}(f).
For other $t/\omega_0$, umklapp scattering immediately gives
rise to long-range order
[Figs.~\ref{fig:lambda6}(c)--\ref{fig:lambda6}(f)],
$K_\rho\to0$
[Fig.~\ref{fig:lambda6}(b)], and insulating behavior [Fig.~\ref{fig:lambda6}(a)].
The clean power-law scaling observed on the BOW-CDW critical line in
Figs.~\ref{fig:lambda6}(c)--\ref{fig:lambda6}(f)
is consistent with the predicted absence
of logarithmic corrections due to the vanishing of the umklapp term
\cite{1989JPhA...22..511A}.
Finally, the absence of a CDW phase in functional RG results for the SSH model
\cite{Ba.Bo.07} may be the result of using a linearized spectrum. This
approximation amounts to an infinite bandwidth $W$, whereas CDW order appears in
Fig.~\ref{fig:phasediagram} at $\omega_0\gtrsim W=4t$.

An intuitive physical picture of how BOW and CDW phases---breaking different
symmetries---can be connected via a generically continuous phase
transition is the scenario of a 1D DQCP~\cite{mudry2019quantum}. It involves
solitons in the CDW (BOW) order parameter that can be added in pairs and interpolate between the two
degenerate CDW (BOW) configurations. Parameterizing the phase of the order
parameter by $\boldsymbol{\varphi}=(\cos\varphi,\sin\varphi)$
\cite{mudry2019quantum}, see the inset of Fig.~\ref{fig:phasediagram}, BOW (CDW)
patterns correspond to $\varphi=0,\pi$ ($\varphi=\pm\pi/2$).  For example, a
defect in the BOW order connecting $\varphi=0,\pi$ contains a region with CDW
order or $\varphi=\pi/2$.  Simultaneous proliferation of BOW/CDW defects at
$(t/\omega_{0})_\text{c}$ provides a mechanism for a continuous transition without
fine-tuning. 

Instead of the bosonized theory~(\ref{eq:bosonized}), the SSH
model~(\ref{eq:SSH0}) can also be described in terms of a Gross-Neveu
field theory of Dirac fermions \cite{PhysRevD.10.3235}. Such a representation
makes topological and symmetry aspects more transparent. While a
Gross-Neveu theory was given in Ref.~\cite{PhysRevB.27.1680}, our numerical
results can only be captured by the more general form with two interactions \cite{PhysRevD.10.3235}
\begin{equation}\label{eq:grossneveu}
  \mathcal{L} = \bpsib\,\rmi\gamma^\mu\partial_\mu\bpsi + g_\text{bow}(\bpsib
  \bpsi)^2+g_\text{cdw}(\bpsib\,\rmi\gamma_5\bpsi)^2\,,
\end{equation}
where $\bpsi=(\psi_{A},\psi_{B})$ ($A$/$B$: sublattices, see Fig.~\ref{fig:phasediagram}).  
Similar to the umklapp term in Eq.~(\ref{eq:bosonized}), the interactions
account for the lattice symmetries that distinguish BOW and CDW order \cite{mudry2014lecture}.
In contrast, a mean-field approximation of Eq.~(\ref{eq:grossneveu}) contains 
mass terms  $g_\text{bow} m_1\bpsib\bpsi$ and $g_\text{cdw} m_2\bpsib\, \rmi
\gamma_5 \bpsi$ \cite{PhysRevLett.47.986}. It can be shown
\cite{fradkin2013field} that these mass terms anticommute and may be rotated
into each other, establishing a chiral U(1) symmetry absent beyond the
mean-field level. Combining the BOW and CDW masses into a vector $\boldsymbol{m}=(m_1,m_2)$,
the mean-field spectrum $E(p)=\pm\sqrt{p^2 +
  |\boldsymbol{m}|^2}$ \cite{PhysRevB.80.205319}. Together with the chiral
symmetry this form reveals the possibility of a continuous evolution between
BOW and CDW order during which $\boldsymbol{m}$ changes its direction from
$(m,0)$ to $(0,m)$ while $|\boldsymbol{m}|=m$ (the mean-field gap) remains
nonzero. The absence of a CDW-BOW phase transition at the mean-field level
reflects the fact that the CDW and BOW order parameters (characterized by
different broken symmetries on the lattice, see Sec.~\ref{Sec:Model}) become
equivalent up to a chiral rotation in the continuum. The different symmetries
are correctly captured by the bosonized theory~(\ref{eq:bosonized}) and also
the Gross-Neveu theory~(\ref{eq:grossneveu}) \cite{mudry2014lecture,fradkin2013field}.
In the absence of a chiral symmetry, the CDW-BOW transition involves a gap
closing, $|\boldsymbol{m}|\to 0$ at the critical point. This agrees
with our numerical results for the spinless SSH model~(\ref{eq:SSH0}), for
which metallic behavior entails a vanishing single-particle gap.

Finally, it is interesting to contrast the BOW-CDW transition considered here
with recent work on interaction-driven QPTs out of a topological {\em band}
insulator \cite{PhysRevB.89.115430,bermudez2018gross,PhysRevB.99.064105} (see
also Ref.~\cite{MHHF2017}). In the latter, a static BOW mass term arises from
a dimerized hopping (the mean-field SSH model). Soliton excitations are 
therefore only possible for the CDW order parameter and the critical behavior
is significantly different. Instead of the deconfined scenario observed here,
the BOW-CDW transition exhibits Ising criticality~\cite{bermudez2018gross}.

\section{Conclusions \& outlook\label{Sec:Conclusions}}

We used an exact QMC method for retarded interactions to determine the
phase diagram of the 1D SSH model with quantum phonons.
In addition to the well-known metallic LL and the Peierls-ordered BOW phase
we found an extended CDW phase at high
phonon frequencies $\omega_0$ that has been absent in previous studies
\cite{PhysRevB.27.1680,PhysRevB.29.4230,Ba.Bo.07,PhysRevB.91.245147}.
We provided evidence that the CDW and BOW ordered phases are connected by a
direct, continuous quantum phase transition with unconventional power-law exponents
$K_\rho < 1/2$ at the metallic critical point. Our findings are consistent
with a bosonized field theory that was originally introduced for the frustrated $XXZ$ chain \cite{PhysRevB.25.4925}.
In analogy with frustrated spin systems \cite{mudry2019quantum}, the CDW-BOW transition can be interpreted in terms
of a 1D DQCP and the proliferation of solitons.

Our results demonstrate that competing orders can be generated from a single retarded interaction
that originates, \eg, from an off-diagonal operator coupled to a phonon.
Instead of having different competing interactions in an equal-time Hamiltonian,
here the interaction range in imaginary time determines whether BOW or CDW order dominates. 
Competing phenomena therefore arise at a critical interaction range determined by the
phonon frequency $\omega_0$. The study of retarded
interactions might be a promising approach to generate complex phase diagrams also
in higher dimensions.
An interesting generalization of the work presented in this paper
is the spinful 2D
SSH model, which supports a phase transition between valence-bond and
antiferromagnetic phases \cite{Be.Ho.Go.As.2019}. Moreover, the relation
between the spinless SSH model and the still incompletely understood
frustrated $XY$ model \cite{PhysRevB.72.184502,HPV-05c,lima2018fully} should be explored.

\begin{acknowledgments}
  We acknowledge helpful discussions with F. Assaad, A. Furusaki, C. Mudry,
  and F. Pollmann. M.W. was supported by the U.S. Department of Energy (DOE),
  Office of Science, Basic Energy Sciences (BES) under Award
  DE-FG02-08ER46542.  F.P.T was funded by the Deutsche Forschungsgemeinschaft
  (DFG, German Research Foundation) -- project number 414456783, M.H. via SFB
  1170. We thank the Gauss Centre for Supercomputing (SuperMUC at the Leibniz
  Supercomputing Centre) for generous allocation of supercomputing resources.
\end{acknowledgments}

\appendix

\section{Quantum Monte Carlo method\label{App:QMC}}

We used the directed-loop QMC method for retarded interactions in the
path-integral representation \cite{arXiv:1704.07913}. It is based on an
interaction expansion of the partition function
$Z= \int \fcohmeasure \, e^{-\S_0 -\S_1}$ around
$\S_0 = \int d\tau \sum_i \fcohcr{i}(\tau) \, \partial_\tau \,
\fcohan{i}(\tau)$.
A general interaction vertex $\S_1 = -\sum_{\nu} w_\nu \, h_\nu$ can be
written as a sum over vertex variables $\nu$, a weight $w_\nu$, and the
Grassmann fields contained in $h_\nu$.  The perturbation expansion becomes
\begin{align}
  \label{eq:perturbation_expansion}
  Z
  =
  \sum_{n=0}^\infty \frac{1}{n!} \sum_{C_n} w_{\nu_1} \dots w_{\nu_n} \int \fcohmeasure \,  e^{-\S_0} h_{\nu_1} \dots h_{\nu_n}
\end{align}
with sums over the expansion order $n$ and the ordered vertex list
$C_n = \{\nu_1,\dots,\nu_n\}$.  For each time-ordered configuration of
vertices, the expectation value over Grassmann fields can be represented by
world lines.  The trivial choice of $\S_0$ ensures that the imaginary-time
evolution is entirely determined by the interaction vertices.  Therefore,
Eq.~(\ref{eq:perturbation_expansion}) is the path-integral equivalent of the
stochastic series expansion (SSE) representation where
$Z=\Tr \, e^{-\beta H}$ is expanded in the total Hamiltonian
\cite{PhysRevB.43.5950,SandvikPhononsa}.  Accordingly, many algorithmic
features, including the global directed-loop updates \cite{Sandvik02},
directly transfer to the path-integral representation \cite{SandvikPhononsa}.

The retarded interaction of the SSH model includes two bond operators acting
at different imaginary times. 
Therefore, a compatible interaction vertex must contain two subvertices
$j\in\{1,2\}$ with local variables $\{a_j,b_j,\tau_j\}$ labeling the operator
type, bond, and time of each operator. For the SSH model with a coupling to
optical bond phonons we have $b_1=b_2=b$. The interaction vertex of the SSH
model becomes
\begin{align}\label{eq:S1vert_spin}
  \S_1 = - \iint_0^\beta d\tau_1 d\tau_2 \, \Pp(\tau_1 - \tau_2)
  \sum_{a_1, a_2, b} h_{a_1a_2,b}(\tau_1,\tau_2) \, .
\end{align}
It is important to note that the symmetrized phonon propagator
$\Pp(\tau) = \omega_0 \cosh[\omega_0 (\beta/2-\tau)] / [2\sinh(\omega_0
\beta/2)]$
is included in the global weight $w_\nu$ of the vertex.  Whereas the
bond-bond interaction
\begin{align}
  \label{eq:vert11}
  h_{11,b}(\tau_1,\tau_2)
  =
  \frac{\lambda t}{2} \, B_{b}(\tau_1) \, B_{b}(\tau_2)
\end{align}
is already nonlocal in time, the single hopping terms of the kinetic energy
are promoted to retarded interactions by including unit operators with a
second time variable, \ie,
\begin{align}
\nonumber
  h_{10,b}(\tau_1,\tau_2)
  &=
  \frac{t}{2} \, B_{b}(\tau_1) \, \mathbb{1}_b (\tau_2) \, \\
  h_{01,b}(\tau_1,\tau_2)
  &=
  \frac{t}{2} \, \mathbb{1}_{b}(\tau_1) \, B_{b}(\tau_2) \, .
   \label{eq:vert10}
\end{align}
This is possible because $\int_0^\beta d\tau_2 \, \Pp(\tau_1-\tau_2) = 1$. As
the vertices (\ref{eq:vert11}) and (\ref{eq:vert10}) both contain
off-diagonal hopping operators, we have to include a purely diagonal term in
the interaction vertex.  The simplest choice is a constant shift of the
action,
\begin{align}
  \label{eq:vert00}
  h_{00,b}(\tau_1,\tau_2)
  = k \,
  \mathbb{1}_{b}(\tau_1) \, \mathbb{1}_{b}(\tau_2) \, .
\end{align}

With our choice of interaction vertices, we can formulate the diagonal and
directed-loop updates similar to the SSE representation \cite{Sandvik02}. For
the diagonal updates, we use the Metropolis algorithm to add and remove
vertices $h_{00,b}(\tau_1,\tau_2)$ that do not change the world-line
configurations but change the expansion order $n$.  We propose time
differences $\tau_1 - \tau_2$ according to the phonon propagator using
inverse-transform sampling.  Because $\Pp(\tau_1 - \tau_2)$ appears as a
global weight in front of each vertex, it drops out of the directed-loop
equations.  The latter can be solved for each vertex similarly to the
original approach, see the Supplemental Material of
Ref.~\cite{arXiv:1704.07913}.  The constant $k$ in Eq.~(\ref{eq:vert00}) has
to be chosen such that every weight in the loop assignments is positive.
During the propagation of the directed loop, unit operators can be
transformed into bond operators and vice versa, leading to local updates
$h_{00,b} \leftrightarrow h_{10,b} / h_{01,b} \leftrightarrow h_{11,b}$. Note
that the vertices are constructed in such a way that each subvertex can be
changed individually while the other subvertex remains unchanged. For details
on the updating schemes, we refer to Refs.~\cite{Sandvik02,arXiv:1704.07913}.

The calculation of observables in the path-integral (interaction)
representation is in many ways similar to the SSE representation.  Sandvik
\etal~\cite{SandvikPhononsa} systematically compared estimators for
electronic correlation functions derived in the two
representations. Estimators that only include diagonal operators, such as the
charge structure factor $C_\rho(r)$ and the charge susceptibility
$\chi_\rho(r)$, are simple to derive and given in
Ref.~\cite{SandvikPhononsa}. Estimators including off-diagonal operators can
often be recovered from the vertex distribution if there is a vertex that
only includes this operator.  Measuring the static or dynamic correlations
functions of two bond operators at arbitrary bonds $b_1$ and $b_2$ is only
possible when considering the hopping vertices $h_{10,b} / h_{01,b}$. It
turns out that the bond susceptibility $\chi_\text{b}(r)$ has a very simple
estimator where only the total number of hopping vertices at bonds $b_1$ or
$b_2$ has to be computed, see Ref.~\cite{SandvikPhononsa} for the exact
estimator.  However, calculating the equal-time bond structure factor
$C_\text{b}(r=b_1-b_2) = \expvtext{B_{b_1} B_{b_2}}$ in the interaction
representation is more involved.  While a general derivation is outlined in
Ref.~\cite{SandvikPhononsa}, we only state the final estimator for the SSH
model. For a Monte Carlo configuration $C_n$, the bond structure factor can
be estimated from
\begin{align}
  C_\text{b}(b_1,b_2;C_n)
  =
  \frac{1}{\beta t^2} \sum_{p} I_{b_1 b_2}(p-1,p) \, K(p-1, p) \, .
\end{align}
In principle, the sum over $p$ runs over the time-ordered list of all
subvertices contained in a world-line configuration.  However, we can exclude
the unit operators $\mathbb{1}_{b}$ as they were only introduced to simplify
the Monte Carlo sampling. $I_{b_1 b_2} (p-1,p)$ is zero unless bond operators
$B_{b_1}(\tau_{p-1})$ and $B_{b_2}(\tau_{p})$ originating from the hopping
terms $h_{10} / h_{01}$ appear at adjacent times; then
$I_{b_1 b_2} (p-1,p)=1$.  An integral expression for $K(p-1,p)$ was derived
in Ref.~\cite{SandvikPhononsa} and gives
$K(p-1,p) = 2/(\tau_{p+1} - \tau_{p-2})$ when 4 or more subvertices are
present in a world-line configuration.  The time difference
$\tau_{p+1} - \tau_{p-2} \in [0,\beta]$ is defined by the two subvertices
that surround the two bond operators under consideration.  Note that
$K(p-1,p)= 2/\beta$ for 3 subvertices, $K(p-1,p) = 1/\beta$ for 2
subvertices, and $K(p-1,p) = 0$ for 0 or 1 subvertices.  For further details,
see Ref.~\cite{SandvikPhononsa}.

The Monte Carlo configurations do not give direct access to observables
containing phonon fields because the latter have been integrated out to
obtain a retarded fermionic interaction. However, bosonic observables can be
recovered from electronic correlation functions using generating
functionals. In particular, we derived efficient estimators for the total
energy, specific heat, fidelity susceptibility, and phonon propagator in
Refs.~\cite{PhysRevB.94.245138, PhysRevB.98.235117} that make use of the
vertex distribution. In the following section, we use the framework outlined
in Ref.~\cite{PhysRevB.94.245138} to show that the superfluid stiffness of an
electron-phonon model can still be calculated from the winding number.

\section{Estimator for the superfluid stiffness\label{App:Stiffness}}

Consider a ring of length $L$ threaded by a magnetic flux $\phi$. At finite
temperatures, the superfluid stiffness can be obtained from the free energy
via \cite{Scalapino93}
\begin{align}
  \rho_\text{s} 
  =
  L \, \frac{\partial^2 F(\phi)}{\partial\phi^2} \bigg\rvert_{\phi=0} \, .
\end{align}
Because we study a 1D system \cite{hetenyi2014drude} and our simulations
at $\beta t=2L$ are essentially converged with respect to temperature, the
measured values of $\rho_\text{s}$ are representative of the charge stiffness
or Drude weight defined in Eq.~(\ref{Eq:Drude}). 

\begin{figure*}
  \includegraphics[width=0.9\textwidth]{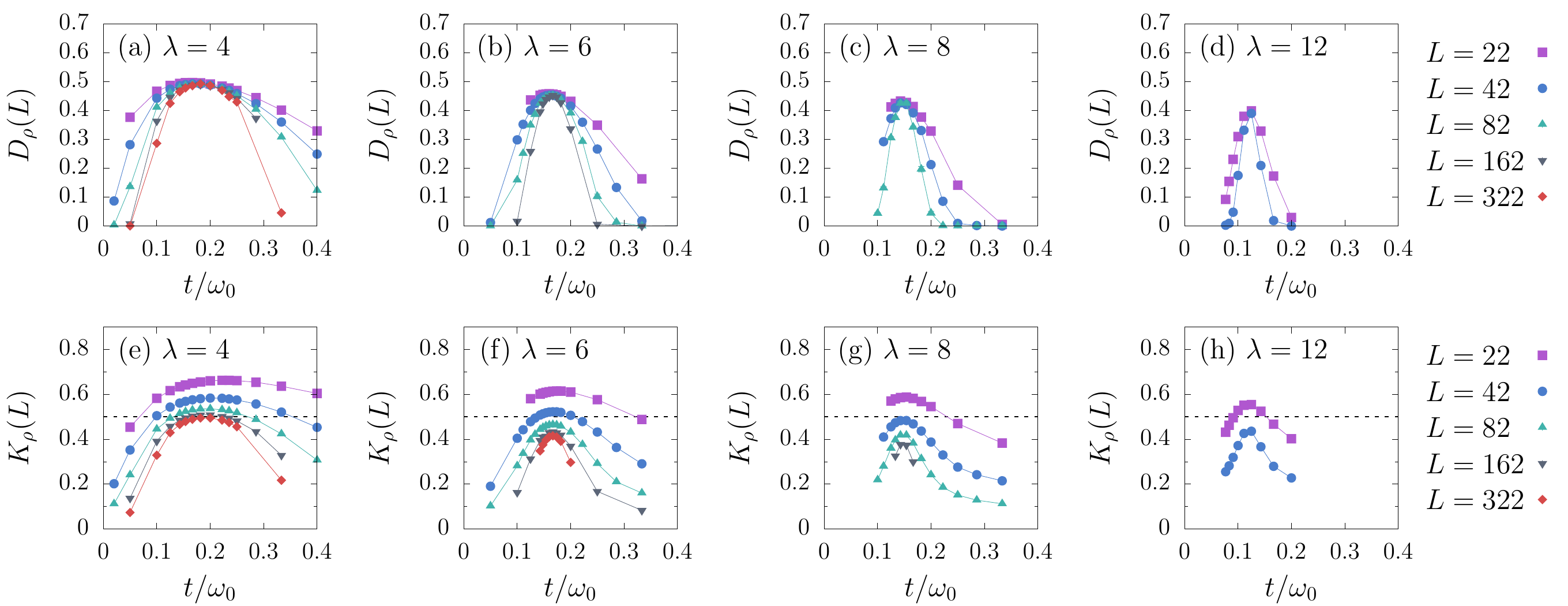}
  \caption{\label{fig:sm:stiffnessvsomega} (a)--(d) Charge stiffness and
    (e)--(h) LL parameter as a function of $t/\omega_0$ for different
    $\lambda$.}
\end{figure*}

Using $F = - \frac{1}{\beta} \ln Z$, the stiffness is directly related to the
action of the SSH model.  The magnetic flux can be incorporated by imposing
twisted boundary conditions $\hat{c}_{L+1} = e^{\im\phi} \hat{c}_1$.  The
boundary term of the action reads
\begin{align}
  \label{eq:Sphi}
  \S_\phi
  =
  \SL \, e^{\im\phi} + \SR \, e^{-\im\phi} + \SLL \, e^{2\im\phi} + \SRR \, e^{-2\im\phi} \, .
\end{align}
Here, $\S_\text{L/R}$ is the action of the hopping term (\ref{eq:vert10})
crossing the boundary to the left/right, whereas $\S_\text{LL/RR}$
corresponds to the bond-bond interaction (\ref{eq:vert11}) with both hopping
operators going to the left/right.  The superfluid stiffness can then be
calculated as
\begin{align}
  \label{eq:rhophi}
  \rho_\text{s} 
  = 
  \frac{L}{\beta} \bigg[ \expv{\frac{\partial\S_\phi}{\partial\phi}}^2
  + \bigg\langle{\frac{\partial^2\S_\phi}{\partial\phi^2}} \bigg\rangle
  - \bigg\langle{\bigg(\frac{\partial\S_\phi}{\partial\phi}\bigg)^2}\bigg\rangle
  \bigg] \bigg\rvert_{\phi=0} \, .
\end{align}
The first expectation value is given by
\begin{align}
\nonumber
  \bigg\langle \frac{\partial\S_\phi}{\partial\phi} \bigg\rangle \bigg\rvert_{\phi=0}
  &=
  \im \expv{\left(\SL - \SR\right) + 2\left(\SLL - \SRR\right)}  \\
  &=
  -\im \expv{\left(\nL - \nR\right) + 2\left(\nLL - \nRR\right)} \, . 
\end{align}
For each Monte Carlo configuration, expectation values of terms $\S_a$
contained in the interaction vertex (\ref{eq:S1vert_spin}) can be obtained by
counting the number of vertices $n_a$ \cite{PhysRevB.94.245138}.  For the
Monte Carlo average we then obtain $\expvtext{\S_a} = - \expvtext{n_a}$. In
the same way, the second term in Eq.~(\ref{eq:rhophi}) becomes
\begin{align}
\nonumber
  \bigg\langle \frac{\partial^2\S_\phi}{\partial\phi^2} \bigg\rangle \bigg\rvert_{\phi=0}
  &=
  - \expv{\left(\SL + \SR\right) + 4 \left(\SLL + \SRR\right)}  \\
  &=
  \expv{\left(\nL + \nR\right) + 4\left(\nLL+\nRR\right)}
  \label{eq:Sphi2}
\end{align}
and the third term is given by
\begin{align}
\nonumber
  \bigg\langle{\bigg(\frac{\partial\S_\phi}{\partial\phi}\bigg)^2}\bigg\rangle \bigg\rvert_{\phi=0}
  =
    &- \expv{\left[\left(\SL-\SR\right)+2\left(\SLL - \SRR\right)\right]^2} \\\nonumber
  =
    &-\expv{\left[ \left(\nL -\nR\right) + 2\left(\nLL - \nRR\right) \right]^2} \\
    &+ \expv{\left(\nL + \nR\right) + 4\left(\nLL + \nRR\right)} \,,
\end{align}
where we used
$\expvtext{\S_a \S_b} = \expvtext{n_a n_b} - \delta_{ab} \expvtext{n_a}$. We
get an additional shift for $a=b$ that cancels the contribution of
(\ref{eq:Sphi2}).  Our results are equivalent to calculating the winding
number $W = n^B_\text{L} - n^B_\text{R}$ where $n^B_\text{L/R}$ counts the
number of subvertices $B_b(\tau)$ crossing the boundary to the
left/right. Here, $n_\text{LL/RR}$ contributes with a factor of 2 because
each vertex contains two bond operators, whereas mixed contributions
$n_\text{LR}$ drop out.  Therefore, $\rho_\text{s}$ can be calculated in the
same way for retarded interactions as for equal-time interactions \cite{PhysRevB.36.8343}, \ie,
\begin{equation}
\rho_s
=
\frac{L}{\beta} (\expvtext{W^2} - \expvtext{W}^2) \, .
\end{equation}

\section{Additional data\label{App:AddData}}

\subsection{CDW-BOW transition\label{App:AddDataOrder}}

Figure~\ref{fig:sm:stiffnessvsomega} shows $D_\rho(L)$ and $K_\rho(L)$ as a
function of $t/\omega_0$ for $\lambda=4,6,8,12$. For all couplings, the data
are consistent with a metallic region at intermediate $t/\omega_0$. Whereas
the apparent narrowing of this region between $\lambda=4$ and $\lambda=6$
matches the phase boundaries in Fig.~\ref{fig:phasediagram}, the theory
discussed in the main text suggests that the BOW-CDW transition involves a
gap closing and hence metallic behavior only at a single point. At this
transition, the LL parameter $K_\rho<1/2$.
Values $K_\rho<1/2$ can be reconciled
with metallic behavior by assuming $\lambda_\phi=0$ in
Eq.~(\ref{eq:bosonized}) at the BOW-CDW critical point
\cite{PhysRevB.25.4925,mudry2014lecture}. Apart
from $\lambda=6$, see Fig.~\ref{fig:lambda6}(b), we also find evidence for
$K_\rho<1/2$ at criticality for $\lambda=4$ [Fig.~\ref{fig:sm:stiffnessvsomega}(e)], consistent with a
location on the BOW-CDW transition line.  Therefore, the two separate
critical points (with significant uncertainty) in
Fig.~\ref{fig:phasediagram}, inferred from
Fig.~\ref{fig:stiffness-criticalvalues}(e),
may be an artifact of the challenging
finite-size scaling in the tricritical region of the phase diagram.

In contrast to Ginzburg-Landau theory, the BOW-CDW transition does not
require fine-tuning of both $t/\omega_0$ and $\lambda$. For a fixed
$\lambda$, $\lambda_\phi$ can be tuned to zero for a suitable value of
$t/\omega_0$, giving rise to a line of critical points. Since $K_\rho<1/2$ at
criticality, any nonzero $\lambda_\phi$ yields long-range BOW or CDW
order. The theory hence excludes an extended metallic region (as opposed to a
critical line) with $K_\rho<1/2$.

Previous work on the extended Hubbard model \cite{Sa.Ba.Ca.04} suggests that
a peak in $K_\rho(L)$ that narrows with increasing $L$ indicates a continuous
transition, whereas the absence of a peak or a broadening with increasing $L$
signals a first-order
transition. Figures~\ref{fig:sm:stiffnessvsomega}(e)--(h) hence support
continuous behavior, in accordance with theoretical
expectations~\cite{PhysRevB.25.4925,mudry2019quantum}.

\subsection{LL-BOW and LL-CDW transition\label{App:AddDataLL}}

\begin{figure}
  \includegraphics[width=\linewidth]{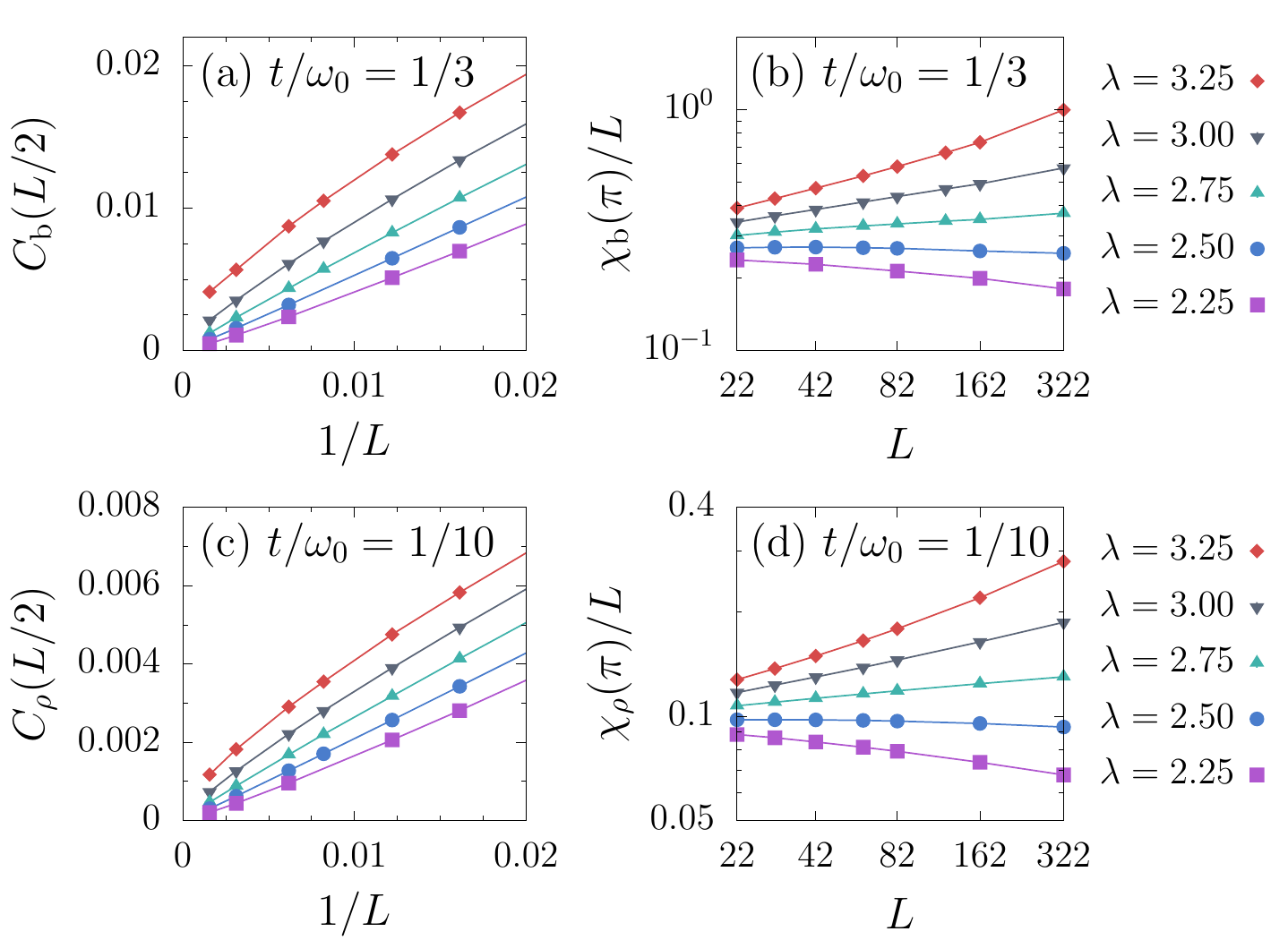}
  \caption{\label{fig:ordersus} 
 (a), (c) Finite-size scaling of the BOW/CDW order parameter $C_{\mathrm{b}/\rho}(L/2)$ and (b), (d) scaling analysis of the susceptibility $\chi_{\mathrm{b}/\rho}(\pi)/L$ at $t/\omega_0=1/3$ and
 $t/\omega_0=1/10$.
    }
\end{figure}

The stiffness fits are based on the characteristic logarithmic scaling at the
critical point. As is commonly done for non-BKT transitions in higher dimensions,
the LL-BOW and LL-CDW critical phase boundaries can also be
estimated from the real-space correlation functions in Eq.~(\ref{eq:correlators}),
evaluated at the maximum distance $L/2$. Results for $C_\mathrm{b}(L/2)$ at
$t/\omega_0=1/3$ and $C_\rho(L/2)$ at $t/\omega_0=1/10$ are shown in
Figs.~\ref{fig:ordersus}(a) and \ref{fig:ordersus}(c), respectively.
For both cases shown, the critical values can be identified
visually as $\lambda_{\mathrm{c}} \approx 2.75$. They agree with our
estimates from the stiffness fits within error bars, but
suggest the possibility that the latter slightly overestimate the
critical couplings. We notice that logarithmic corrections are expected to appear
close to the transition.
A similar analysis was done for the $t$-$V$ model in Ref.~\cite{PhysRevB.92.245132}.

The critical points of the LL-BOW and LL-CDW phase transitions can further be
deduced from the bond and charge susceptibilities defined in Eq.~(\ref{Eq:Sus}). In the LL phase,
$\chi_{\mathrm{b}/\rho}(\pi)/L \sim L^{1-\overline{\eta}}$ with $\overline{\eta} = 2K_\rho$.
If we assume $K_\rho = 1/2$ at the critical point---as predicted
theoretically ($K_\rho<1/2$ is only expected at BOW-CDW critical points \cite{PhysRevB.25.4925})---$\chi_{\mathrm{b}/\rho}(\pi)/L$
will converge to a constant at $\lambda = \lambda_\mathrm{c}$ but scale to zero
(diverge) for $\lambda < \lambda_\mathrm{c}$ ($\lambda > \lambda_\mathrm{c}$). In contrast to the CDW-BOW transition considered
in Sec.~\ref{Sec:CDW-BOW}, the presence of logarithmic corrections close
to the critical point complicates a precise estimation of $\lambda_\mathrm{c}$.
It was recently demonstrated for the $t$-$V$ model
that $\chi_{\mathrm{b}/\rho}(\pi)/L$ at $\lambda=\lambda_\mathrm{c}$
increases slowly as a function of $L$
for the system sizes considered here \cite{PhysRevB.92.245132}.
Figures~\ref{fig:ordersus}(b) and \ref{fig:ordersus}(d) are therefore in agreement with
our estimates $\lambda_\mathrm{c} \approx 2.75$ from
$C_{\mathrm{b}/\rho}(L/2)$. In principle, critical values can also be
extracted from the correlation length, which shows a logarithmic
scaling similar to the stiffness at $\lambda=\lambda_c$ \cite{PhysRevE.87.032105}.

The critical values from Fig.~\ref{fig:ordersus} are compatible with
those from the stiffness fits. The assumption of
$K_\rho = 1/2$ at $\lambda_\mathrm{c}$ can give a more precise estimate of
$\lambda_\mathrm{c}$ when using the susceptibility instead of the stiffness fits.
However, the stiffness fits are more general and do not require $K_\rho = 1/2$.
The fact that all our estimates are consistent
does not imply $K_\rho = 1/2$ at $\lambda_\mathrm{c}$.
On the other hand, we do not find evidence for $K_\rho < 1/2$
at the LL-BOW and LL-CDW transitions, contrary to predictions from
functional RG calculations for the LL-BOW transition in the SSH model \cite{Ba.Bo.07}.

\section{Stiffness fits\label{App:StiffnessFits}}

\begin{figure}
  \includegraphics[width=0.45\textwidth]{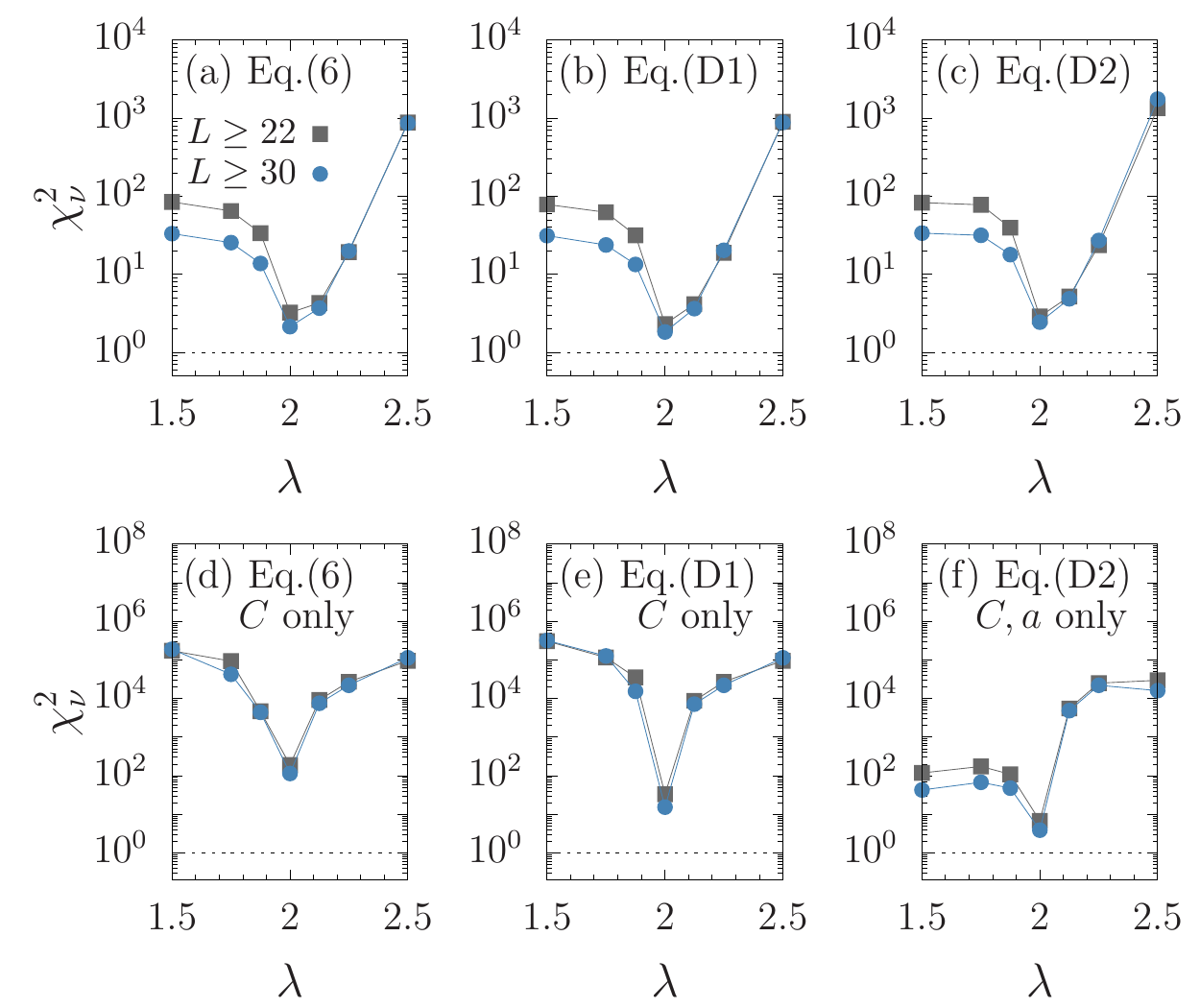}
  \caption{\label{fig:fits:ominf} Comparison of different fit functions and
    minimal system sizes for $t/\omega_0=0$ (corresponding to the $t$-$V$
    model). The exact critical value is $\lambda_c=2$. The bottom row shows
    fits based on $D_\rho(\infty)=t/2$ and $g=1$.
    }
\end{figure}

Standard BKT universality is predicted for the LL-BOW and LL-CDW transitions
both in a general LL \cite{Giamarchi} and specifically for the frustrated $XXZ$
chain \cite{PhysRevB.25.4925}. A detailed RG analysis
\cite{PhysRevE.87.032105} gives the finite-size scaling forms
\begin{align}\label{eq:fit2-1}
  \frac{D_\rho(L)}{D_\rho(\infty)}
  = 
    1 &+ \frac{g}{2\ln L + C + \ln (C/2 + \ln L)}\,,\\\label{eq:fit2-2}
  \frac{D_\rho(L)}{D_\rho(\infty)}
  = 
    1 &+ \frac{g}{2\ln L + C + \ln (C/2 + \ln L)} \\
    &+ \frac{a}{[2\ln L + C + \ln (C/2 + \ln L)]^2}\,,
    \nonumber
\end{align}
which provide the leading corrections to Eq.~(\ref{eq:WeberMinnhagen}).
However, in light of the observed nonuniversal jumps, functional RG
predictions of $K_\rho<1/2$ at the LL-BOW transition \cite{Ba.Bo.07}, and
$K_\rho<1/2$ at the BOW-CDW transition according to our data and theory
\cite{PhysRevB.25.4925}, we determined the critical values in
Fig.~\ref{fig:phasediagram} using fits based on Eq.~(\ref{eq:WeberMinnhagen})
with three parameters: $D_\rho(\infty)$, $g$, and $C$.  In contrast, $g$ and
$D_\rho(\infty)$ can be computed exactly for the classical 2D $XY$ model (see
below), leaving only one free parameter. Specifically, for
$\beta t\sim L_y=\infty$ (1D quantum chain at $T=0$), $g=1$ and
$D_\rho=2/\pi$ ($D_\rho=t/2$) for the 2D $XY$ (1D $t$-$V$) model
\cite{hasenbusch2005two,laflorencie2001finite}.  As expected and demonstrated
below, multi-parameter fits provide less accurate, but nonetheless fully
consistent, critical values (shallower minima, stronger dependence on the
range of $L$) than single-parameter fits. This is particularly relevant for
the analysis of quantum systems such as the SSH model, where the range and
number of system sizes are limited.

For a fit involving $N$ data points, \ie, stiffness values $O_n$ for
different system sizes $L_n\in\{L_1,L_2,\dots,L_N\}$ with corresponding statistical errors $\sigma_n$, 
the reduced $\chi^2$ is calculated from 
\begin{equation}
  \chi^2_\nu = \frac{1}{\nu}\chi^2 = \frac{1}{\nu}\sum_{n=1}^N \left(\frac{O_n - C_n}{\sigma_n}\right)^2\,.
\end{equation}
Here, the number of degrees of freedom $\nu$ is given by $N-M$, where $M$ is the
number of fit parameters, and $C_n$ is the stiffness value predicted by the
fit for system size $L_n$.

For the fits, we restricted the range of the jump to
$0<D_\rho(\infty)<{2t}/{\pi}$, using the known value of the noninteracting
case. To discriminate between the logarithmic scaling at the critical point
and the very weak finite-size dependence at weak coupling [see
Fig.~\ref{fig:stiffness-criticalvalues}(a)], a nonzero lower bound
$g_\text{min}$ was imposed. Otherwise, the choice $g=0$ gives good fits
throughout the LL phase and there would be no minimum of $\chi^2_\nu$ at the
critical point. The exact value of $g_\text{min}$ does not significantly
affect the results and was chosen as $0.25$.  Finally, the allowed range of
$C$ was $[0,\infty[$.

An important test case for the generalized, multi-parameter fit
ansatz~(\ref{eq:WeberMinnhagen}) was the LL-CDW transition of the $t$-$V$
model, for which the critical value is known. We used the same range of
system sizes as for the SSH model. Figures~\ref{fig:fits:ominf}(a)--(c) give
a comparison of results based on
Eqs.~(\ref{eq:WeberMinnhagen}),~(\ref{eq:fit2-1}), and~(\ref{eq:fit2-2}).
All three fit functions yield very similar and hence compatible minima of
$\chi^2_\nu$ at the correct value
$\lambda=2$. Figures~\ref{fig:fits:ominf}(d)--(f) are based on fits that
exploit the known values $g=1$ and $D_\rho(\infty)=t/2$.  This additional
information produces significantly sharper minima, in accordance with
previous work on 2D $XY$ models \cite{PhysRevB.37.5986}.  At the same time, the
first-order fit functions~(\ref{eq:WeberMinnhagen}) and~(\ref{eq:fit2-1}) do
not fully capture the finite-size scaling on small system sizes, as
manifested in $\chi^2_\nu\gg 1$ even at $\lambda=2$ in
Figs.~\ref{fig:fits:ominf}(d) and (e) for $L\geq 22$ and $L\geq
30$.
Higher-order corrections are partially captured by varying $g$ and
$D_\rho(\infty)$ \cite{hasenbusch2005two}, which explains the much better
$\chi^2_\nu$ for the same range of $L$ in Figs.~\ref{fig:fits:ominf}(a) and
(b).

\begin{figure}[b]
  \includegraphics[width=0.45\textwidth]{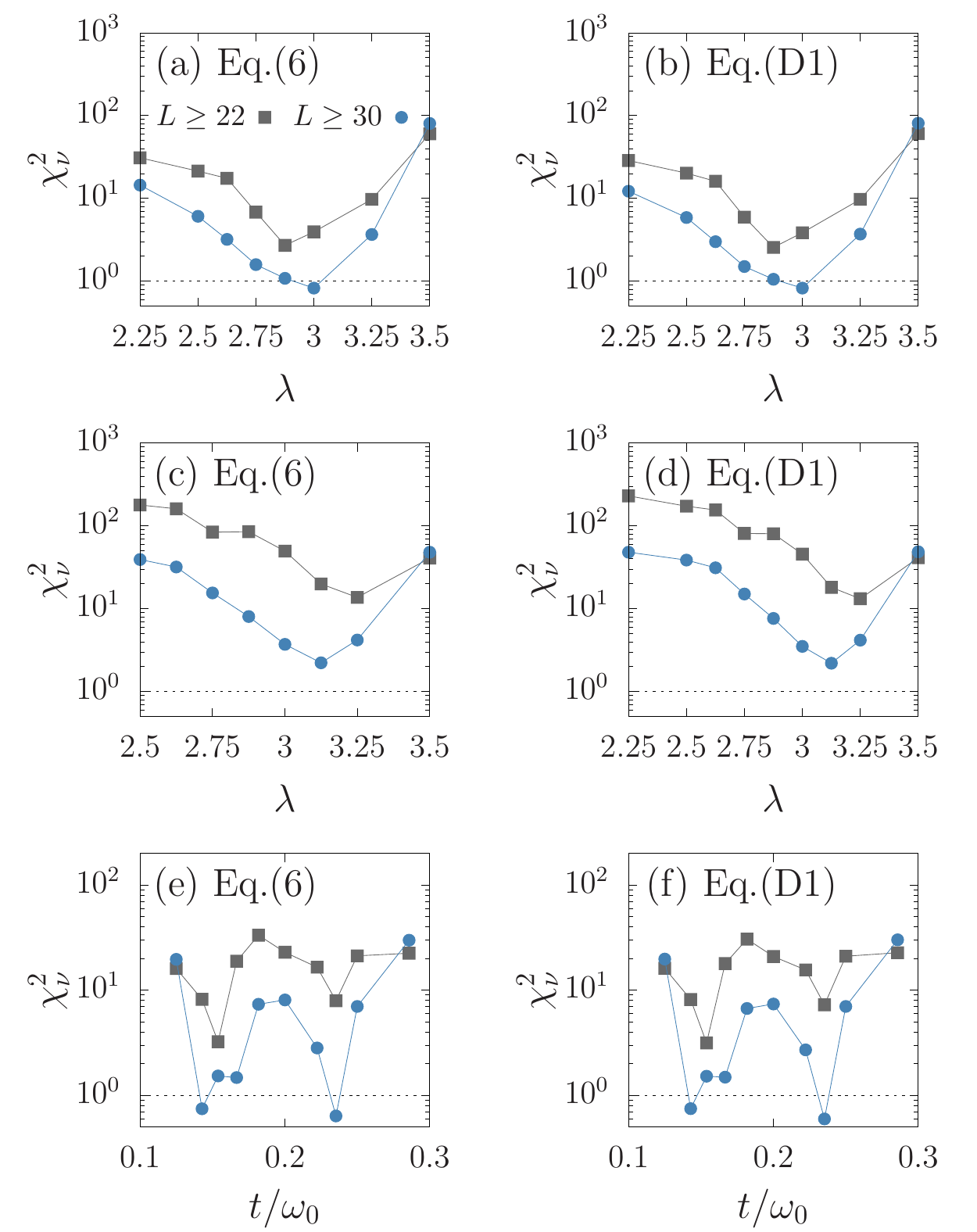}
  \caption{\label{fig:fits:om10} Stiffness fits for (a),(b)
    $t/\omega_0=1/10$, (c),(d) $t/\omega_0=1/3$, and (e),(f) $\lambda=4$. 
    }
\end{figure}

For the more challenging case of $t/\omega_0>0$, we focused on
three-parameter fits based on Eqs.~(\ref{eq:WeberMinnhagen}) and
(\ref{eq:fit2-1}). Within the present accuracy, the results
in Fig.~\ref{fig:fits:om10}
 are
compatible with each other but slightly less systematic than for the $t$-$V$
model. In particular, the fits become less robust upon increasing the
smallest $L$ due to a reduced number of degrees of freedom. A
similar picture arises for a fixed $\lambda=4$ in
Figs.~\ref{fig:fits:om10}(e) and \ref{fig:fits:om10}(f).
For the present accuracy and range of
system sizes, we cannot discriminate between the scaling
forms~(\ref{eq:WeberMinnhagen}),~(\ref{eq:fit2-1}), and~(\ref{eq:fit2-2}).

\section{Nonuniversal stiffness jumps\label{App:StiffnessJumps}}

For the $XY$ model, the stiffness jump and the constant $g$ can be computed
from a series for a given aspect ratio $r=L_x/L_y$
\cite{hasenbusch2005two,PhysRevB.69.014509}. For 1+1D quantum systems,
$r=cL/\beta$, with $c$ a model-dependent constant. For example,
$D_\rho(\infty)$ varies significantly as a function of $L_x/L_y$, covering
the whole range from $2/\pi$ to $0$ \cite{PhysRevB.69.014509}. Similarly, $g$
varies between $1$ and $\infty$ as a function of $r$. In principle, a change
of the range of the retarded interaction can mimic a change in $r$, leading
to a dependence of $D_\rho(\infty)$ and $g$ on the phonon frequency.

There are several other known mechanisms for nonuniversal values of the
stiffness jump. The bosonization relation $D_\rho=K_\rho u$ \cite{Giamarchi}
implies that, even if $K_\rho=1/2$ at a QPT, $D_\rho(\infty)$ can change via
the renormalized velocity $u$. For example, $u$ increases with $V$ in the
$t$-$V$ model \cite{shankar1990solvable} but decreases with $\lambda$ in the
Holstein model \cite{PhysRevB.98.235117}. The stiffness can also be reduced
by non-vortex excitations that are not captured by the standard BKT theory of
the $XY$ model \cite{BKTsinegordon}.

%\bibliography{../refs}

\begin{thebibliography}{87}%
\makeatletter
\providecommand \@ifxundefined [1]{%
 \@ifx{#1\undefined}
}%
\providecommand \@ifnum [1]{%
 \ifnum #1\expandafter \@firstoftwo
 \else \expandafter \@secondoftwo
 \fi
}%
\providecommand \@ifx [1]{%
 \ifx #1\expandafter \@firstoftwo
 \else \expandafter \@secondoftwo
 \fi
}%
\providecommand \natexlab [1]{#1}%
\providecommand \enquote  [1]{``#1''}%
\providecommand \bibnamefont  [1]{#1}%
\providecommand \bibfnamefont [1]{#1}%
\providecommand \citenamefont [1]{#1}%
\providecommand \href@noop [0]{\@secondoftwo}%
\providecommand \href [0]{\begingroup \@sanitize@url \@href}%
\providecommand \@href[1]{\@@startlink{#1}\@@href}%
\providecommand \@@href[1]{\endgroup#1\@@endlink}%
\providecommand \@sanitize@url [0]{\catcode `\\12\catcode `\$12\catcode
  `\&12\catcode `\#12\catcode `\^12\catcode `\_12\catcode `\%12\relax}%
\providecommand \@@startlink[1]{}%
\providecommand \@@endlink[0]{}%
\providecommand \url  [0]{\begingroup\@sanitize@url \@url }%
\providecommand \@url [1]{\endgroup\@href {#1}{\urlprefix }}%
\providecommand \urlprefix  [0]{URL }%
\providecommand \Eprint [0]{\href }%
\providecommand \doibase [0]{http://dx.doi.org/}%
\providecommand \selectlanguage [0]{\@gobble}%
\providecommand \bibinfo  [0]{\@secondoftwo}%
\providecommand \bibfield  [0]{\@secondoftwo}%
\providecommand \translation [1]{[#1]}%
\providecommand \BibitemOpen [0]{}%
\providecommand \bibitemStop [0]{}%
\providecommand \bibitemNoStop [0]{.\EOS\space}%
\providecommand \EOS [0]{\spacefactor3000\relax}%
\providecommand \BibitemShut  [1]{\csname bibitem#1\endcsname}%
\let\auto@bib@innerbib\@empty
%</preamble>
\bibitem [{\citenamefont {Manzeli}\ \emph {et~al.}(2017)\citenamefont
  {Manzeli}, \citenamefont {Ovchinnikov}, \citenamefont {Pasquier},
  \citenamefont {Yazyev},\ and\ \citenamefont {Kis}}]{manzeli20172d}%
  \BibitemOpen
  \bibfield  {author} {\bibinfo {author} {\bibfnamefont {S.}~\bibnamefont
  {Manzeli}}, \bibinfo {author} {\bibfnamefont {D.}~\bibnamefont
  {Ovchinnikov}}, \bibinfo {author} {\bibfnamefont {D.}~\bibnamefont
  {Pasquier}}, \bibinfo {author} {\bibfnamefont {O.~V.}\ \bibnamefont
  {Yazyev}}, \ and\ \bibinfo {author} {\bibfnamefont {A.}~\bibnamefont {Kis}},\
  }\href@noop {} {\bibfield  {journal} {\bibinfo  {journal} {Nat. Rev.
  Materials}\ }\textbf {\bibinfo {volume} {2}},\ \bibinfo {pages} {17033}
  (\bibinfo {year} {2017})}\BibitemShut {NoStop}%
\bibitem [{\citenamefont {Lee}\ \emph {et~al.}(2006)\citenamefont {Lee},
  \citenamefont {Nagaosa},\ and\ \citenamefont {Wen}}]{RevModPhys.78.17}%
  \BibitemOpen
  \bibfield  {author} {\bibinfo {author} {\bibfnamefont {P.~A.}\ \bibnamefont
  {Lee}}, \bibinfo {author} {\bibfnamefont {N.}~\bibnamefont {Nagaosa}}, \ and\
  \bibinfo {author} {\bibfnamefont {X.-G.}\ \bibnamefont {Wen}},\ }\href
  {\doibase 10.1103/RevModPhys.78.17} {\bibfield  {journal} {\bibinfo
  {journal} {Rev. Mod. Phys.}\ }\textbf {\bibinfo {volume} {78}},\ \bibinfo
  {pages} {17} (\bibinfo {year} {2006})}\BibitemShut {NoStop}%
\bibitem [{\citenamefont {Sato}\ \emph {et~al.}(2017)\citenamefont {Sato},
  \citenamefont {Hohenadler},\ and\ \citenamefont {Assaad}}]{sato2017dirac}%
  \BibitemOpen
  \bibfield  {author} {\bibinfo {author} {\bibfnamefont {T.}~\bibnamefont
  {Sato}}, \bibinfo {author} {\bibfnamefont {M.}~\bibnamefont {Hohenadler}}, \
  and\ \bibinfo {author} {\bibfnamefont {F.~F.}\ \bibnamefont {Assaad}},\
  }\href {\doibase 10.1103/PhysRevLett.119.197203} {\bibfield  {journal}
  {\bibinfo  {journal} {Phys. Rev. Lett.}\ }\textbf {\bibinfo {volume} {119}},\
  \bibinfo {pages} {197203} (\bibinfo {year} {2017})}\BibitemShut {NoStop}%
\bibitem [{\citenamefont {Motruk}\ \emph {et~al.}(2015)\citenamefont {Motruk},
  \citenamefont {Grushin}, \citenamefont {de~Juan},\ and\ \citenamefont
  {Pollmann}}]{PhysRevB.92.085147}%
  \BibitemOpen
  \bibfield  {author} {\bibinfo {author} {\bibfnamefont {J.}~\bibnamefont
  {Motruk}}, \bibinfo {author} {\bibfnamefont {A.~G.}\ \bibnamefont {Grushin}},
  \bibinfo {author} {\bibfnamefont {F.}~\bibnamefont {de~Juan}}, \ and\
  \bibinfo {author} {\bibfnamefont {F.}~\bibnamefont {Pollmann}},\ }\href
  {\doibase 10.1103/PhysRevB.92.085147} {\bibfield  {journal} {\bibinfo
  {journal} {Phys. Rev. B}\ }\textbf {\bibinfo {volume} {92}},\ \bibinfo
  {pages} {085147} (\bibinfo {year} {2015})}\BibitemShut {NoStop}%
\bibitem [{\citenamefont {Li}\ \emph {et~al.}(2017)\citenamefont {Li},
  \citenamefont {Jiang}, \citenamefont {Jian},\ and\ \citenamefont
  {Yao}}]{li2017fermion}%
  \BibitemOpen
  \bibfield  {author} {\bibinfo {author} {\bibfnamefont {Z.-X.}\ \bibnamefont
  {Li}}, \bibinfo {author} {\bibfnamefont {Y.-F.}\ \bibnamefont {Jiang}},
  \bibinfo {author} {\bibfnamefont {S.-K.}\ \bibnamefont {Jian}}, \ and\
  \bibinfo {author} {\bibfnamefont {H.}~\bibnamefont {Yao}},\ }\href@noop {}
  {\bibfield  {journal} {\bibinfo  {journal} {Nature Comm.}\ }\textbf {\bibinfo
  {volume} {8}},\ \bibinfo {pages} {314} (\bibinfo {year} {2017})}\BibitemShut
  {NoStop}%
\bibitem [{\citenamefont {de~la Pe\~na}\ \emph {et~al.}(2017)\citenamefont
  {de~la Pe\~na}, \citenamefont {Lichtenstein},\ and\ \citenamefont
  {Honerkamp}}]{PhysRevB.95.085143}%
  \BibitemOpen
  \bibfield  {author} {\bibinfo {author} {\bibfnamefont {D.~S.}\ \bibnamefont
  {de~la Pe\~na}}, \bibinfo {author} {\bibfnamefont {J.}~\bibnamefont
  {Lichtenstein}}, \ and\ \bibinfo {author} {\bibfnamefont {C.}~\bibnamefont
  {Honerkamp}},\ }\href {\doibase 10.1103/PhysRevB.95.085143} {\bibfield
  {journal} {\bibinfo  {journal} {Phys. Rev. B}\ }\textbf {\bibinfo {volume}
  {95}},\ \bibinfo {pages} {085143} (\bibinfo {year} {2017})}\BibitemShut
  {NoStop}%
\bibitem [{\citenamefont {Janssen}\ \emph {et~al.}(2018)\citenamefont
  {Janssen}, \citenamefont {Herbut},\ and\ \citenamefont
  {Scherer}}]{PhysRevB.97.041117}%
  \BibitemOpen
  \bibfield  {author} {\bibinfo {author} {\bibfnamefont {L.}~\bibnamefont
  {Janssen}}, \bibinfo {author} {\bibfnamefont {I.~F.}\ \bibnamefont {Herbut}},
  \ and\ \bibinfo {author} {\bibfnamefont {M.~M.}\ \bibnamefont {Scherer}},\
  }\href {\doibase 10.1103/PhysRevB.97.041117} {\bibfield  {journal} {\bibinfo
  {journal} {Phys. Rev. B}\ }\textbf {\bibinfo {volume} {97}},\ \bibinfo
  {pages} {041117} (\bibinfo {year} {2018})}\BibitemShut {NoStop}%
\bibitem [{\citenamefont {He}\ \emph {et~al.}(2018)\citenamefont {He},
  \citenamefont {Xu}, \citenamefont {Sun}, \citenamefont {Assaad},
  \citenamefont {Meng},\ and\ \citenamefont {Lu}}]{PhysRevB.97.081110}%
  \BibitemOpen
  \bibfield  {author} {\bibinfo {author} {\bibfnamefont {Y.-Y.}\ \bibnamefont
  {He}}, \bibinfo {author} {\bibfnamefont {X.~Y.}\ \bibnamefont {Xu}}, \bibinfo
  {author} {\bibfnamefont {K.}~\bibnamefont {Sun}}, \bibinfo {author}
  {\bibfnamefont {F.~F.}\ \bibnamefont {Assaad}}, \bibinfo {author}
  {\bibfnamefont {Z.~Y.}\ \bibnamefont {Meng}}, \ and\ \bibinfo {author}
  {\bibfnamefont {Z.-Y.}\ \bibnamefont {Lu}},\ }\href {\doibase
  10.1103/PhysRevB.97.081110} {\bibfield  {journal} {\bibinfo  {journal} {Phys.
  Rev. B}\ }\textbf {\bibinfo {volume} {97}},\ \bibinfo {pages} {081110}
  (\bibinfo {year} {2018})}\BibitemShut {NoStop}%
\bibitem [{\citenamefont {Liu}\ \emph {et~al.}(2019)\citenamefont {Liu},
  \citenamefont {Wang}, \citenamefont {Sato}, \citenamefont {Hohenadler},
  \citenamefont {Wang}, \citenamefont {Guo},\ and\ \citenamefont
  {Assaad}}]{liu2019superconductivity}%
  \BibitemOpen
  \bibfield  {author} {\bibinfo {author} {\bibfnamefont {Y.}~\bibnamefont
  {Liu}}, \bibinfo {author} {\bibfnamefont {Z.}~\bibnamefont {Wang}}, \bibinfo
  {author} {\bibfnamefont {T.}~\bibnamefont {Sato}}, \bibinfo {author}
  {\bibfnamefont {M.}~\bibnamefont {Hohenadler}}, \bibinfo {author}
  {\bibfnamefont {C.}~\bibnamefont {Wang}}, \bibinfo {author} {\bibfnamefont
  {W.}~\bibnamefont {Guo}}, \ and\ \bibinfo {author} {\bibfnamefont {F.~F.}\
  \bibnamefont {Assaad}},\ }\href@noop {} {\bibfield  {journal} {\bibinfo
  {journal} {Nature communications}\ }\textbf {\bibinfo {volume} {10}},\
  \bibinfo {pages} {2658} (\bibinfo {year} {2019})}\BibitemShut {NoStop}%
\bibitem [{\citenamefont {Senthil}\ \emph {et~al.}(2004)\citenamefont
  {Senthil}, \citenamefont {Vishwanath}, \citenamefont {Balents}, \citenamefont
  {Sachdev},\ and\ \citenamefont {Fisher}}]{senthil2004decon}%
  \BibitemOpen
  \bibfield  {author} {\bibinfo {author} {\bibfnamefont {T.}~\bibnamefont
  {Senthil}}, \bibinfo {author} {\bibfnamefont {A.}~\bibnamefont {Vishwanath}},
  \bibinfo {author} {\bibfnamefont {L.}~\bibnamefont {Balents}}, \bibinfo
  {author} {\bibfnamefont {S.}~\bibnamefont {Sachdev}}, \ and\ \bibinfo
  {author} {\bibfnamefont {M.~P.~A.}\ \bibnamefont {Fisher}},\ }\href {\doibase
  10.1126/science.1091806} {\bibfield  {journal} {\bibinfo  {journal}
  {Science}\ }\textbf {\bibinfo {volume} {303}},\ \bibinfo {pages} {1490}
  (\bibinfo {year} {2004})}\BibitemShut {NoStop}%
\bibitem [{\citenamefont {Nahum}\ \emph {et~al.}(2015)\citenamefont {Nahum},
  \citenamefont {Chalker}, \citenamefont {Serna}, \citenamefont {Ortu\~no},\
  and\ \citenamefont {Somoza}}]{Nahum15}%
  \BibitemOpen
  \bibfield  {author} {\bibinfo {author} {\bibfnamefont {A.}~\bibnamefont
  {Nahum}}, \bibinfo {author} {\bibfnamefont {J.~T.}\ \bibnamefont {Chalker}},
  \bibinfo {author} {\bibfnamefont {P.}~\bibnamefont {Serna}}, \bibinfo
  {author} {\bibfnamefont {M.}~\bibnamefont {Ortu\~no}}, \ and\ \bibinfo
  {author} {\bibfnamefont {A.~M.}\ \bibnamefont {Somoza}},\ }\href {\doibase
  10.1103/PhysRevX.5.041048} {\bibfield  {journal} {\bibinfo  {journal} {Phys.
  Rev. X}\ }\textbf {\bibinfo {volume} {5}},\ \bibinfo {pages} {041048}
  (\bibinfo {year} {2015})}\BibitemShut {NoStop}%
\bibitem [{\citenamefont {Shao}\ \emph {et~al.}(2016)\citenamefont {Shao},
  \citenamefont {Guo},\ and\ \citenamefont {Sandvik}}]{Shao15}%
  \BibitemOpen
  \bibfield  {author} {\bibinfo {author} {\bibfnamefont {H.}~\bibnamefont
  {Shao}}, \bibinfo {author} {\bibfnamefont {W.}~\bibnamefont {Guo}}, \ and\
  \bibinfo {author} {\bibfnamefont {A.~W.}\ \bibnamefont {Sandvik}},\ }\href
  {\doibase 10.1126/science.aad5007} {\bibfield  {journal} {\bibinfo  {journal}
  {Science}\ }\textbf {\bibinfo {volume} {352}},\ \bibinfo {pages} {213}
  (\bibinfo {year} {2016})}\BibitemShut {NoStop}%
\bibitem [{\citenamefont {Wang}\ \emph {et~al.}(2017)\citenamefont {Wang},
  \citenamefont {Nahum}, \citenamefont {Metlitski}, \citenamefont {Xu},\ and\
  \citenamefont {Senthil}}]{wang2017decon}%
  \BibitemOpen
  \bibfield  {author} {\bibinfo {author} {\bibfnamefont {C.}~\bibnamefont
  {Wang}}, \bibinfo {author} {\bibfnamefont {A.}~\bibnamefont {Nahum}},
  \bibinfo {author} {\bibfnamefont {M.~A.}\ \bibnamefont {Metlitski}}, \bibinfo
  {author} {\bibfnamefont {C.}~\bibnamefont {Xu}}, \ and\ \bibinfo {author}
  {\bibfnamefont {T.}~\bibnamefont {Senthil}},\ }\href {\doibase
  10.1103/PhysRevX.7.031051} {\bibfield  {journal} {\bibinfo  {journal} {Phys.
  Rev. X}\ }\textbf {\bibinfo {volume} {7}},\ \bibinfo {pages} {031051}
  (\bibinfo {year} {2017})}\BibitemShut {NoStop}%
\bibitem [{\citenamefont {Qin}\ \emph {et~al.}(2017)\citenamefont {Qin},
  \citenamefont {He}, \citenamefont {You}, \citenamefont {Lu}, \citenamefont
  {Sen}, \citenamefont {Sandvik}, \citenamefont {Xu},\ and\ \citenamefont
  {Meng}}]{qin2017dual}%
  \BibitemOpen
  \bibfield  {author} {\bibinfo {author} {\bibfnamefont {Y.~Q.}\ \bibnamefont
  {Qin}}, \bibinfo {author} {\bibfnamefont {Y.-Y.}\ \bibnamefont {He}},
  \bibinfo {author} {\bibfnamefont {Y.-Z.}\ \bibnamefont {You}}, \bibinfo
  {author} {\bibfnamefont {Z.-Y.}\ \bibnamefont {Lu}}, \bibinfo {author}
  {\bibfnamefont {A.}~\bibnamefont {Sen}}, \bibinfo {author} {\bibfnamefont
  {A.~W.}\ \bibnamefont {Sandvik}}, \bibinfo {author} {\bibfnamefont
  {C.}~\bibnamefont {Xu}}, \ and\ \bibinfo {author} {\bibfnamefont {Z.~Y.}\
  \bibnamefont {Meng}},\ }\href {\doibase 10.1103/PhysRevX.7.031052} {\bibfield
   {journal} {\bibinfo  {journal} {Phys. Rev. X}\ }\textbf {\bibinfo {volume}
  {7}},\ \bibinfo {pages} {031052} (\bibinfo {year} {2017})}\BibitemShut
  {NoStop}%
\bibitem [{\citenamefont {Jiang}\ and\ \citenamefont
  {Motrunich}(2019)}]{PhysRevB.99.075103}%
  \BibitemOpen
  \bibfield  {author} {\bibinfo {author} {\bibfnamefont {S.}~\bibnamefont
  {Jiang}}\ and\ \bibinfo {author} {\bibfnamefont {O.}~\bibnamefont
  {Motrunich}},\ }\href {\doibase 10.1103/PhysRevB.99.075103} {\bibfield
  {journal} {\bibinfo  {journal} {Phys. Rev. B}\ }\textbf {\bibinfo {volume}
  {99}},\ \bibinfo {pages} {075103} (\bibinfo {year} {2019})}\BibitemShut
  {NoStop}%
\bibitem [{\citenamefont {Mudry}\ \emph {et~al.}(2019)\citenamefont {Mudry},
  \citenamefont {Furusaki}, \citenamefont {Morimoto},\ and\ \citenamefont
  {Hikihara}}]{mudry2019quantum}%
  \BibitemOpen
  \bibfield  {author} {\bibinfo {author} {\bibfnamefont {C.}~\bibnamefont
  {Mudry}}, \bibinfo {author} {\bibfnamefont {A.}~\bibnamefont {Furusaki}},
  \bibinfo {author} {\bibfnamefont {T.}~\bibnamefont {Morimoto}}, \ and\
  \bibinfo {author} {\bibfnamefont {T.}~\bibnamefont {Hikihara}},\ }\href@noop
  {} {\bibfield  {journal} {\bibinfo  {journal} {Physical Review B}\ }\textbf
  {\bibinfo {volume} {99}},\ \bibinfo {pages} {205153} (\bibinfo {year}
  {2019})}\BibitemShut {NoStop}%
\bibitem [{\citenamefont {Roberts}\ \emph {et~al.}(2019)\citenamefont
  {Roberts}, \citenamefont {Jiang},\ and\ \citenamefont
  {Motrunich}}]{1DdcqpRoberts}%
  \BibitemOpen
  \bibfield  {author} {\bibinfo {author} {\bibfnamefont {B.}~\bibnamefont
  {Roberts}}, \bibinfo {author} {\bibfnamefont {S.}~\bibnamefont {Jiang}}, \
  and\ \bibinfo {author} {\bibfnamefont {O.~I.}\ \bibnamefont {Motrunich}},\
  }\href@noop {} {\bibfield  {journal} {\bibinfo  {journal} {Physical Review
  B}\ }\textbf {\bibinfo {volume} {99}},\ \bibinfo {pages} {165143} (\bibinfo
  {year} {2019})}\BibitemShut {NoStop}%
\bibitem [{\citenamefont {Huang}\ \emph {et~al.}(2019)\citenamefont {Huang},
  \citenamefont {Lu}, \citenamefont {You}, \citenamefont {Meng},\ and\
  \citenamefont {Xiang}}]{1DdqcpHuang}%
  \BibitemOpen
  \bibfield  {author} {\bibinfo {author} {\bibfnamefont {R.-Z.}\ \bibnamefont
  {Huang}}, \bibinfo {author} {\bibfnamefont {D.-C.}\ \bibnamefont {Lu}},
  \bibinfo {author} {\bibfnamefont {Y.-Z.}\ \bibnamefont {You}}, \bibinfo
  {author} {\bibfnamefont {Z.~Y.}\ \bibnamefont {Meng}}, \ and\ \bibinfo
  {author} {\bibfnamefont {T.}~\bibnamefont {Xiang}},\ }\href {\doibase
  10.1103/PhysRevB.100.125137} {\bibfield  {journal} {\bibinfo  {journal}
  {Phys. Rev. B}\ }\textbf {\bibinfo {volume} {100}},\ \bibinfo {pages}
  {125137} (\bibinfo {year} {2019})}\BibitemShut {NoStop}%
\bibitem [{\citenamefont {Raghu}\ \emph {et~al.}(2008)\citenamefont {Raghu},
  \citenamefont {Qi}, \citenamefont {Honerkamp},\ and\ \citenamefont
  {Zhang}}]{Raghu08}%
  \BibitemOpen
  \bibfield  {author} {\bibinfo {author} {\bibfnamefont {S.}~\bibnamefont
  {Raghu}}, \bibinfo {author} {\bibfnamefont {X.-L.}\ \bibnamefont {Qi}},
  \bibinfo {author} {\bibfnamefont {C.}~\bibnamefont {Honerkamp}}, \ and\
  \bibinfo {author} {\bibfnamefont {S.-C.}\ \bibnamefont {Zhang}},\ }\href
  {\doibase 10.1103/PhysRevLett.100.156401} {\bibfield  {journal} {\bibinfo
  {journal} {Phys. Rev. Lett.}\ }\textbf {\bibinfo {volume} {100}},\ \bibinfo
  {eid} {156401} (\bibinfo {year} {2008})}\BibitemShut {NoStop}%
\bibitem [{\citenamefont {Hohenadler}\ and\ \citenamefont
  {Fehske}(2018)}]{MHHF2017}%
  \BibitemOpen
  \bibfield  {author} {\bibinfo {author} {\bibfnamefont {M.}~\bibnamefont
  {Hohenadler}}\ and\ \bibinfo {author} {\bibfnamefont {H.}~\bibnamefont
  {Fehske}},\ }\href@noop {} {\bibfield  {journal} {\bibinfo  {journal} {Eur.
  Phys. J. B}\ }\textbf {\bibinfo {volume} {91}},\ \bibinfo {pages} {204}
  (\bibinfo {year} {2018})}\BibitemShut {NoStop}%
\bibitem [{\citenamefont {Capponi}(2016)}]{capponi2016phase}%
  \BibitemOpen
  \bibfield  {author} {\bibinfo {author} {\bibfnamefont {S.}~\bibnamefont
  {Capponi}},\ }\href@noop {} {\bibfield  {journal} {\bibinfo  {journal} {J.
  Phys.: Condens. Matter}\ }\textbf {\bibinfo {volume} {29}},\ \bibinfo {pages}
  {043002} (\bibinfo {year} {2016})}\BibitemShut {NoStop}%
\bibitem [{\citenamefont {Bardeen}\ \emph {et~al.}(1957)\citenamefont
  {Bardeen}, \citenamefont {Cooper},\ and\ \citenamefont
  {Schrieffer}}]{PhysRev.108.1175}%
  \BibitemOpen
  \bibfield  {author} {\bibinfo {author} {\bibfnamefont {J.}~\bibnamefont
  {Bardeen}}, \bibinfo {author} {\bibfnamefont {L.~N.}\ \bibnamefont {Cooper}},
  \ and\ \bibinfo {author} {\bibfnamefont {J.~R.}\ \bibnamefont {Schrieffer}},\
  }\href {\doibase 10.1103/PhysRev.108.1175} {\bibfield  {journal} {\bibinfo
  {journal} {Phys. Rev.}\ }\textbf {\bibinfo {volume} {108}},\ \bibinfo {pages}
  {1175} (\bibinfo {year} {1957})}\BibitemShut {NoStop}%
\bibitem [{\citenamefont {Pouget}(2016)}]{Pouget2016332}%
  \BibitemOpen
  \bibfield  {author} {\bibinfo {author} {\bibfnamefont {J.-P.}\ \bibnamefont
  {Pouget}},\ }\href@noop {} {\bibfield  {journal} {\bibinfo  {journal}
  {Comptes Rendus Physique}\ }\textbf {\bibinfo {volume} {17}},\ \bibinfo
  {pages} {332 } (\bibinfo {year} {2016})}\BibitemShut {NoStop}%
\bibitem [{\citenamefont {Su}\ \emph {et~al.}(1979)\citenamefont {Su},
  \citenamefont {Schrieffer},\ and\ \citenamefont
  {Heeger}}]{PhysRevLett.42.1698}%
  \BibitemOpen
  \bibfield  {author} {\bibinfo {author} {\bibfnamefont {W.~P.}\ \bibnamefont
  {Su}}, \bibinfo {author} {\bibfnamefont {J.~R.}\ \bibnamefont {Schrieffer}},
  \ and\ \bibinfo {author} {\bibfnamefont {A.~J.}\ \bibnamefont {Heeger}},\
  }\href {\doibase 10.1103/PhysRevLett.42.1698} {\bibfield  {journal} {\bibinfo
   {journal} {Phys. Rev. Lett.}\ }\textbf {\bibinfo {volume} {42}},\ \bibinfo
  {pages} {1698} (\bibinfo {year} {1979})}\BibitemShut {NoStop}%
\bibitem [{\citenamefont {Fradkin}(2013)}]{fradkin2013field}%
  \BibitemOpen
  \bibfield  {author} {\bibinfo {author} {\bibfnamefont {E.}~\bibnamefont
  {Fradkin}},\ }\href@noop {} {\emph {\bibinfo {title} {Field theories of
  condensed matter physics}}}\ (\bibinfo  {publisher} {Cambridge University
  Press},\ \bibinfo {year} {2013})\BibitemShut {NoStop}%
\bibitem [{\citenamefont {Verresen}\ \emph {et~al.}(2017)\citenamefont
  {Verresen}, \citenamefont {Moessner},\ and\ \citenamefont
  {Pollmann}}]{PhysRevB.96.165124}%
  \BibitemOpen
  \bibfield  {author} {\bibinfo {author} {\bibfnamefont {R.}~\bibnamefont
  {Verresen}}, \bibinfo {author} {\bibfnamefont {R.}~\bibnamefont {Moessner}},
  \ and\ \bibinfo {author} {\bibfnamefont {F.}~\bibnamefont {Pollmann}},\
  }\href {\doibase 10.1103/PhysRevB.96.165124} {\bibfield  {journal} {\bibinfo
  {journal} {Phys. Rev. B}\ }\textbf {\bibinfo {volume} {96}},\ \bibinfo
  {pages} {165124} (\bibinfo {year} {2017})}\BibitemShut {NoStop}%
\bibitem [{\citenamefont {Yoshida}\ \emph {et~al.}(2014)\citenamefont
  {Yoshida}, \citenamefont {Peters}, \citenamefont {Fujimoto},\ and\
  \citenamefont {Kawakami}}]{PhysRevLett.112.196404}%
  \BibitemOpen
  \bibfield  {author} {\bibinfo {author} {\bibfnamefont {T.}~\bibnamefont
  {Yoshida}}, \bibinfo {author} {\bibfnamefont {R.}~\bibnamefont {Peters}},
  \bibinfo {author} {\bibfnamefont {S.}~\bibnamefont {Fujimoto}}, \ and\
  \bibinfo {author} {\bibfnamefont {N.}~\bibnamefont {Kawakami}},\ }\href
  {\doibase 10.1103/PhysRevLett.112.196404} {\bibfield  {journal} {\bibinfo
  {journal} {Phys. Rev. Lett.}\ }\textbf {\bibinfo {volume} {112}},\ \bibinfo
  {pages} {196404} (\bibinfo {year} {2014})}\BibitemShut {NoStop}%
\bibitem [{\citenamefont {McGinley}\ and\ \citenamefont
  {Cooper}(2018)}]{PhysRevLett.121.090401}%
  \BibitemOpen
  \bibfield  {author} {\bibinfo {author} {\bibfnamefont {M.}~\bibnamefont
  {McGinley}}\ and\ \bibinfo {author} {\bibfnamefont {N.~R.}\ \bibnamefont
  {Cooper}},\ }\href {\doibase 10.1103/PhysRevLett.121.090401} {\bibfield
  {journal} {\bibinfo  {journal} {Phys. Rev. Lett.}\ }\textbf {\bibinfo
  {volume} {121}},\ \bibinfo {pages} {090401} (\bibinfo {year}
  {2018})}\BibitemShut {NoStop}%
\bibitem [{\citenamefont {Porta}\ \emph {et~al.}(2018)\citenamefont {Porta},
  \citenamefont {Ziani}, \citenamefont {Kennes}, \citenamefont {Gambetta},
  \citenamefont {Sassetti},\ and\ \citenamefont
  {Cavaliere}}]{PhysRevB.98.214306}%
  \BibitemOpen
  \bibfield  {author} {\bibinfo {author} {\bibfnamefont {S.}~\bibnamefont
  {Porta}}, \bibinfo {author} {\bibfnamefont {N.~T.}\ \bibnamefont {Ziani}},
  \bibinfo {author} {\bibfnamefont {D.~M.}\ \bibnamefont {Kennes}}, \bibinfo
  {author} {\bibfnamefont {F.~M.}\ \bibnamefont {Gambetta}}, \bibinfo {author}
  {\bibfnamefont {M.}~\bibnamefont {Sassetti}}, \ and\ \bibinfo {author}
  {\bibfnamefont {F.}~\bibnamefont {Cavaliere}},\ }\href {\doibase
  10.1103/PhysRevB.98.214306} {\bibfield  {journal} {\bibinfo  {journal} {Phys.
  Rev. B}\ }\textbf {\bibinfo {volume} {98}},\ \bibinfo {pages} {214306}
  (\bibinfo {year} {2018})}\BibitemShut {NoStop}%
\bibitem [{\citenamefont {Cirac}\ \emph {et~al.}(2010)\citenamefont {Cirac},
  \citenamefont {Maraner},\ and\ \citenamefont
  {Pachos}}]{PhysRevLett.105.190403}%
  \BibitemOpen
  \bibfield  {author} {\bibinfo {author} {\bibfnamefont {J.~I.}\ \bibnamefont
  {Cirac}}, \bibinfo {author} {\bibfnamefont {P.}~\bibnamefont {Maraner}}, \
  and\ \bibinfo {author} {\bibfnamefont {J.~K.}\ \bibnamefont {Pachos}},\
  }\href {\doibase 10.1103/PhysRevLett.105.190403} {\bibfield  {journal}
  {\bibinfo  {journal} {Phys. Rev. Lett.}\ }\textbf {\bibinfo {volume} {105}},\
  \bibinfo {pages} {190403} (\bibinfo {year} {2010})}\BibitemShut {NoStop}%
\bibitem [{\citenamefont {Bermudez}\ \emph {et~al.}(2018)\citenamefont
  {Bermudez}, \citenamefont {Tirrito}, \citenamefont {Rizzi}, \citenamefont
  {Lewenstein},\ and\ \citenamefont {Hands}}]{bermudez2018gross}%
  \BibitemOpen
  \bibfield  {author} {\bibinfo {author} {\bibfnamefont {A.}~\bibnamefont
  {Bermudez}}, \bibinfo {author} {\bibfnamefont {E.}~\bibnamefont {Tirrito}},
  \bibinfo {author} {\bibfnamefont {M.}~\bibnamefont {Rizzi}}, \bibinfo
  {author} {\bibfnamefont {M.}~\bibnamefont {Lewenstein}}, \ and\ \bibinfo
  {author} {\bibfnamefont {S.}~\bibnamefont {Hands}},\ }\href@noop {}
  {\bibfield  {journal} {\bibinfo  {journal} {Annal. Phys.}\ }\textbf {\bibinfo
  {volume} {399}},\ \bibinfo {pages} {149} (\bibinfo {year}
  {2018})}\BibitemShut {NoStop}%
\bibitem [{\citenamefont {Kuno}\ \emph {et~al.}(2018)\citenamefont {Kuno},
  \citenamefont {Ichinose},\ and\ \citenamefont
  {Takahashi}}]{kuno2018generalized}%
  \BibitemOpen
  \bibfield  {author} {\bibinfo {author} {\bibfnamefont {Y.}~\bibnamefont
  {Kuno}}, \bibinfo {author} {\bibfnamefont {I.}~\bibnamefont {Ichinose}}, \
  and\ \bibinfo {author} {\bibfnamefont {Y.}~\bibnamefont {Takahashi}},\
  }\href@noop {} {\bibfield  {journal} {\bibinfo  {journal} {Sci. Rep.}\
  }\textbf {\bibinfo {volume} {8}},\ \bibinfo {pages} {10699} (\bibinfo {year}
  {2018})}\BibitemShut {NoStop}%
\bibitem [{\citenamefont {Atala}\ \emph {et~al.}(2013)\citenamefont {Atala},
  \citenamefont {Aidelsburger}, \citenamefont {Barreiro}, \citenamefont
  {Abanin}, \citenamefont {Kitagawa}, \citenamefont {Demler},\ and\
  \citenamefont {Bloch}}]{atala2013direct}%
  \BibitemOpen
  \bibfield  {author} {\bibinfo {author} {\bibfnamefont {M.}~\bibnamefont
  {Atala}}, \bibinfo {author} {\bibfnamefont {M.}~\bibnamefont {Aidelsburger}},
  \bibinfo {author} {\bibfnamefont {J.~T.}\ \bibnamefont {Barreiro}}, \bibinfo
  {author} {\bibfnamefont {D.}~\bibnamefont {Abanin}}, \bibinfo {author}
  {\bibfnamefont {T.}~\bibnamefont {Kitagawa}}, \bibinfo {author}
  {\bibfnamefont {E.}~\bibnamefont {Demler}}, \ and\ \bibinfo {author}
  {\bibfnamefont {I.}~\bibnamefont {Bloch}},\ }\href@noop {} {\bibfield
  {journal} {\bibinfo  {journal} {Nature Physics}\ }\textbf {\bibinfo {volume}
  {9}},\ \bibinfo {pages} {795} (\bibinfo {year} {2013})}\BibitemShut {NoStop}%
\bibitem [{\citenamefont {Lee}\ \emph {et~al.}(2018)\citenamefont {Lee},
  \citenamefont {Imhof}, \citenamefont {Berger}, \citenamefont {Bayer},
  \citenamefont {Brehm}, \citenamefont {Molenkamp}, \citenamefont {Kiessling},\
  and\ \citenamefont {Thomale}}]{lee2018topolectrical}%
  \BibitemOpen
  \bibfield  {author} {\bibinfo {author} {\bibfnamefont {C.~H.}\ \bibnamefont
  {Lee}}, \bibinfo {author} {\bibfnamefont {S.}~\bibnamefont {Imhof}}, \bibinfo
  {author} {\bibfnamefont {C.}~\bibnamefont {Berger}}, \bibinfo {author}
  {\bibfnamefont {F.}~\bibnamefont {Bayer}}, \bibinfo {author} {\bibfnamefont
  {J.}~\bibnamefont {Brehm}}, \bibinfo {author} {\bibfnamefont {L.~W.}\
  \bibnamefont {Molenkamp}}, \bibinfo {author} {\bibfnamefont {T.}~\bibnamefont
  {Kiessling}}, \ and\ \bibinfo {author} {\bibfnamefont {R.}~\bibnamefont
  {Thomale}},\ }\href@noop {} {\bibfield  {journal} {\bibinfo  {journal}
  {Communications Physics}\ }\textbf {\bibinfo {volume} {1}},\ \bibinfo {pages}
  {39} (\bibinfo {year} {2018})}\BibitemShut {NoStop}%
\bibitem [{\citenamefont {Fradkin}\ and\ \citenamefont
  {Hirsch}(1983)}]{PhysRevB.27.1680}%
  \BibitemOpen
  \bibfield  {author} {\bibinfo {author} {\bibfnamefont {E.}~\bibnamefont
  {Fradkin}}\ and\ \bibinfo {author} {\bibfnamefont {J.~E.}\ \bibnamefont
  {Hirsch}},\ }\href {\doibase 10.1103/PhysRevB.27.1680} {\bibfield  {journal}
  {\bibinfo  {journal} {Phys. Rev. B}\ }\textbf {\bibinfo {volume} {27}},\
  \bibinfo {pages} {1680} (\bibinfo {year} {1983})}\BibitemShut {NoStop}%
\bibitem [{\citenamefont {Haldane}(1982)}]{PhysRevB.25.4925}%
  \BibitemOpen
  \bibfield  {author} {\bibinfo {author} {\bibfnamefont {F.~D.~M.}\
  \bibnamefont {Haldane}},\ }\href {\doibase 10.1103/PhysRevB.25.4925}
  {\bibfield  {journal} {\bibinfo  {journal} {Phys. Rev. B}\ }\textbf {\bibinfo
  {volume} {25}},\ \bibinfo {pages} {4925} (\bibinfo {year}
  {1982})}\BibitemShut {NoStop}%
\bibitem [{\citenamefont {Sengupta}\ \emph {et~al.}(2003)\citenamefont
  {Sengupta}, \citenamefont {Sandvik},\ and\ \citenamefont
  {Campbell}}]{PhysRevB.67.245103}%
  \BibitemOpen
  \bibfield  {author} {\bibinfo {author} {\bibfnamefont {P.}~\bibnamefont
  {Sengupta}}, \bibinfo {author} {\bibfnamefont {A.~W.}\ \bibnamefont
  {Sandvik}}, \ and\ \bibinfo {author} {\bibfnamefont {D.~K.}\ \bibnamefont
  {Campbell}},\ }\href {\doibase 10.1103/PhysRevB.67.245103} {\bibfield
  {journal} {\bibinfo  {journal} {Phys. Rev. B}\ }\textbf {\bibinfo {volume}
  {67}},\ \bibinfo {pages} {245103} (\bibinfo {year} {2003})}\BibitemShut
  {NoStop}%
\bibitem [{\citenamefont {Weber}\ \emph {et~al.}(2015)\citenamefont {Weber},
  \citenamefont {Assaad},\ and\ \citenamefont
  {Hohenadler}}]{PhysRevB.91.245147}%
  \BibitemOpen
  \bibfield  {author} {\bibinfo {author} {\bibfnamefont {M.}~\bibnamefont
  {Weber}}, \bibinfo {author} {\bibfnamefont {F.~F.}\ \bibnamefont {Assaad}}, \
  and\ \bibinfo {author} {\bibfnamefont {M.}~\bibnamefont {Hohenadler}},\
  }\href {\doibase 10.1103/PhysRevB.91.245147} {\bibfield  {journal} {\bibinfo
  {journal} {Phys. Rev. B}\ }\textbf {\bibinfo {volume} {91}},\ \bibinfo
  {pages} {245147} (\bibinfo {year} {2015})}\BibitemShut {NoStop}%
\bibitem [{\citenamefont {Bakrim}\ and\ \citenamefont
  {Bourbonnais}(2015)}]{Bakrim2015}%
  \BibitemOpen
  \bibfield  {author} {\bibinfo {author} {\bibfnamefont {H.}~\bibnamefont
  {Bakrim}}\ and\ \bibinfo {author} {\bibfnamefont {C.}~\bibnamefont
  {Bourbonnais}},\ }\href {\doibase 10.1103/PhysRevB.91.085114} {\bibfield
  {journal} {\bibinfo  {journal} {Phys. Rev. B}\ }\textbf {\bibinfo {volume}
  {91}},\ \bibinfo {pages} {085114} (\bibinfo {year} {2015})}\BibitemShut
  {NoStop}%
\bibitem [{\citenamefont {Peierls}(1979)}]{Peierls}%
  \BibitemOpen
  \bibfield  {author} {\bibinfo {author} {\bibfnamefont {R.}~\bibnamefont
  {Peierls}},\ }\href@noop {} {\emph {\bibinfo {title} {Surprises in
  Theoretical Physics}}}\ (\bibinfo  {publisher} {Princeton University Press},\
  \bibinfo {address} {New Jersey},\ \bibinfo {year} {1979})\BibitemShut
  {NoStop}%
\bibitem [{\citenamefont {Bakrim}\ and\ \citenamefont
  {Bourbonnais}(2007)}]{Ba.Bo.07}%
  \BibitemOpen
  \bibfield  {author} {\bibinfo {author} {\bibfnamefont {H.}~\bibnamefont
  {Bakrim}}\ and\ \bibinfo {author} {\bibfnamefont {C.}~\bibnamefont
  {Bourbonnais}},\ }\href@noop {} {\bibfield  {journal} {\bibinfo  {journal}
  {Phys. Rev. B}\ }\textbf {\bibinfo {volume} {76}},\ \bibinfo {pages} {195115}
  (\bibinfo {year} {2007})}\BibitemShut {NoStop}%
\bibitem [{\citenamefont {Shankar}(1990)}]{shankar1990solvable}%
  \BibitemOpen
  \bibfield  {author} {\bibinfo {author} {\bibfnamefont {R.}~\bibnamefont
  {Shankar}},\ }\href@noop {} {\bibfield  {journal} {\bibinfo  {journal} {Int.
  J. Mod. Phys. B}\ }\textbf {\bibinfo {volume} {4}},\ \bibinfo {pages} {2371}
  (\bibinfo {year} {1990})}\BibitemShut {NoStop}%
\bibitem [{\citenamefont {Caron}\ and\ \citenamefont
  {Bourbonnais}(1984)}]{PhysRevB.29.4230}%
  \BibitemOpen
  \bibfield  {author} {\bibinfo {author} {\bibfnamefont {L.~G.}\ \bibnamefont
  {Caron}}\ and\ \bibinfo {author} {\bibfnamefont {C.}~\bibnamefont
  {Bourbonnais}},\ }\href {\doibase 10.1103/PhysRevB.29.4230} {\bibfield
  {journal} {\bibinfo  {journal} {Phys. Rev. B}\ }\textbf {\bibinfo {volume}
  {29}},\ \bibinfo {pages} {4230} (\bibinfo {year} {1984})}\BibitemShut
  {NoStop}%
\bibitem [{\citenamefont {Caron}\ and\ \citenamefont
  {Moukouri}(1996)}]{PhysRevLett.76.4050}%
  \BibitemOpen
  \bibfield  {author} {\bibinfo {author} {\bibfnamefont {L.~G.}\ \bibnamefont
  {Caron}}\ and\ \bibinfo {author} {\bibfnamefont {S.}~\bibnamefont
  {Moukouri}},\ }\href {\doibase 10.1103/PhysRevLett.76.4050} {\bibfield
  {journal} {\bibinfo  {journal} {Phys. Rev. Lett.}\ }\textbf {\bibinfo
  {volume} {76}},\ \bibinfo {pages} {4050} (\bibinfo {year}
  {1996})}\BibitemShut {NoStop}%
\bibitem [{\citenamefont {Zheng}(1997)}]{PhysRevB.56.14414}%
  \BibitemOpen
  \bibfield  {author} {\bibinfo {author} {\bibfnamefont {H.}~\bibnamefont
  {Zheng}},\ }\href {\doibase 10.1103/PhysRevB.56.14414} {\bibfield  {journal}
  {\bibinfo  {journal} {Phys. Rev. B}\ }\textbf {\bibinfo {volume} {56}},\
  \bibinfo {pages} {14414} (\bibinfo {year} {1997})}\BibitemShut {NoStop}%
\bibitem [{\citenamefont {Citro}\ \emph {et~al.}(2005)\citenamefont {Citro},
  \citenamefont {Orignac},\ and\ \citenamefont {Giamarchi}}]{Ci.Or.Gi.05}%
  \BibitemOpen
  \bibfield  {author} {\bibinfo {author} {\bibfnamefont {R.}~\bibnamefont
  {Citro}}, \bibinfo {author} {\bibfnamefont {E.}~\bibnamefont {Orignac}}, \
  and\ \bibinfo {author} {\bibfnamefont {T.}~\bibnamefont {Giamarchi}},\
  }\href@noop {} {\bibfield  {journal} {\bibinfo  {journal} {Phys. Rev. B.}\
  }\textbf {\bibinfo {volume} {72}},\ \bibinfo {pages} {024434} (\bibinfo
  {year} {2005})}\BibitemShut {NoStop}%
\bibitem [{\citenamefont {Sil}(1998)}]{sil1998spin}%
  \BibitemOpen
  \bibfield  {author} {\bibinfo {author} {\bibfnamefont {S.}~\bibnamefont
  {Sil}},\ }\href@noop {} {\bibfield  {journal} {\bibinfo  {journal} {J. Phys.:
  Condens. Matter}\ }\textbf {\bibinfo {volume} {10}},\ \bibinfo {pages} {8851}
  (\bibinfo {year} {1998})}\BibitemShut {NoStop}%
\bibitem [{\citenamefont {Sandvik}\ and\ \citenamefont
  {Kurkij{\"a}rvi}(1991)}]{PhysRevB.43.5950}%
  \BibitemOpen
  \bibfield  {author} {\bibinfo {author} {\bibfnamefont {A.~W.}\ \bibnamefont
  {Sandvik}}\ and\ \bibinfo {author} {\bibfnamefont {J.}~\bibnamefont
  {Kurkij{\"a}rvi}},\ }\href@noop {} {\bibfield  {journal} {\bibinfo  {journal}
  {Phys. Rev. B}\ }\textbf {\bibinfo {volume} {43}},\ \bibinfo {pages} {5950}
  (\bibinfo {year} {1991})}\BibitemShut {NoStop}%
\bibitem [{\citenamefont {Syljuasen}\ and\ \citenamefont
  {Sandvik}(2002)}]{Sandvik02}%
  \BibitemOpen
  \bibfield  {author} {\bibinfo {author} {\bibfnamefont {O.}~\bibnamefont
  {Syljuasen}}\ and\ \bibinfo {author} {\bibfnamefont {A.~W.}\ \bibnamefont
  {Sandvik}},\ }\href@noop {} {\bibfield  {journal} {\bibinfo  {journal} {Phys.
  Rev. E}\ }\textbf {\bibinfo {volume} {66}},\ \bibinfo {pages} {046701}
  (\bibinfo {year} {2002})}\BibitemShut {NoStop}%
\bibitem [{\citenamefont {Weber}\ \emph {et~al.}(2017)\citenamefont {Weber},
  \citenamefont {Assaad},\ and\ \citenamefont {Hohenadler}}]{arXiv:1704.07913}%
  \BibitemOpen
  \bibfield  {author} {\bibinfo {author} {\bibfnamefont {M.}~\bibnamefont
  {Weber}}, \bibinfo {author} {\bibfnamefont {F.~F.}\ \bibnamefont {Assaad}}, \
  and\ \bibinfo {author} {\bibfnamefont {M.}~\bibnamefont {Hohenadler}},\
  }\href {\doibase 10.1103/PhysRevLett.119.097401} {\bibfield  {journal}
  {\bibinfo  {journal} {Phys. Rev. Lett.}\ }\textbf {\bibinfo {volume} {119}},\
  \bibinfo {pages} {097401} (\bibinfo {year} {2017})}\BibitemShut {NoStop}%
\bibitem [{\citenamefont {Giamarchi}(2004)}]{Giamarchi}%
  \BibitemOpen
  \bibfield  {author} {\bibinfo {author} {\bibfnamefont {T.}~\bibnamefont
  {Giamarchi}},\ }\href@noop {} {\emph {\bibinfo {title} {Quantum {P}hysics in
  {O}ne {D}imension}}}\ (\bibinfo  {publisher} {Clarendon Press},\ \bibinfo
  {address} {Oxford},\ \bibinfo {year} {2004})\BibitemShut {NoStop}%
\bibitem [{\citenamefont {Kohn}(1964)}]{PhysRev.133.A171}%
  \BibitemOpen
  \bibfield  {author} {\bibinfo {author} {\bibfnamefont {W.}~\bibnamefont
  {Kohn}},\ }\href {\doibase 10.1103/PhysRev.133.A171} {\bibfield  {journal}
  {\bibinfo  {journal} {Phys. Rev.}\ }\textbf {\bibinfo {volume} {133}},\
  \bibinfo {pages} {A171} (\bibinfo {year} {1964})}\BibitemShut {NoStop}%
\bibitem [{\citenamefont {Scalapino}\ \emph {et~al.}(1993)\citenamefont
  {Scalapino}, \citenamefont {White},\ and\ \citenamefont
  {Zhang}}]{Scalapino93}%
  \BibitemOpen
  \bibfield  {author} {\bibinfo {author} {\bibfnamefont {D.~J.}\ \bibnamefont
  {Scalapino}}, \bibinfo {author} {\bibfnamefont {S.}~\bibnamefont {White}}, \
  and\ \bibinfo {author} {\bibfnamefont {S.}~\bibnamefont {Zhang}},\
  }\href@noop {} {\bibfield  {journal} {\bibinfo  {journal} {Phys. Rev. B}\
  }\textbf {\bibinfo {volume} {47}},\ \bibinfo {pages} {7995} (\bibinfo {year}
  {1993})}\BibitemShut {NoStop}%
\bibitem [{\citenamefont {Sengupta}\ \emph {et~al.}(2002)\citenamefont
  {Sengupta}, \citenamefont {Sandvik},\ and\ \citenamefont
  {Campbell}}]{PhysRevB.65.155113}%
  \BibitemOpen
  \bibfield  {author} {\bibinfo {author} {\bibfnamefont {P.}~\bibnamefont
  {Sengupta}}, \bibinfo {author} {\bibfnamefont {A.~W.}\ \bibnamefont
  {Sandvik}}, \ and\ \bibinfo {author} {\bibfnamefont {D.~K.}\ \bibnamefont
  {Campbell}},\ }\href {\doibase 10.1103/PhysRevB.65.155113} {\bibfield
  {journal} {\bibinfo  {journal} {Phys. Rev. B}\ }\textbf {\bibinfo {volume}
  {65}},\ \bibinfo {pages} {155113} (\bibinfo {year} {2002})}\BibitemShut
  {NoStop}%
\bibitem [{\citenamefont {Hasenbusch}\ \emph
  {et~al.}(2005{\natexlab{a}})\citenamefont {Hasenbusch}, \citenamefont
  {Pelissetto},\ and\ \citenamefont {Vicari}}]{HPV-05c}%
  \BibitemOpen
  \bibfield  {author} {\bibinfo {author} {\bibfnamefont {M.}~\bibnamefont
  {Hasenbusch}}, \bibinfo {author} {\bibfnamefont {A.}~\bibnamefont
  {Pelissetto}}, \ and\ \bibinfo {author} {\bibfnamefont {E.}~\bibnamefont
  {Vicari}},\ }\href {\doibase 10.1088/1742-5468/2005/12/P12002} {\bibfield
  {journal} {\bibinfo  {journal} {J.~Stat.~Mech.}\ }\textbf {\bibinfo {volume}
  {12}},\ \bibinfo {pages} {12002} (\bibinfo {year}
  {2005}{\natexlab{a}})}\BibitemShut {NoStop}%
\bibitem [{\citenamefont {Kosterlitz}(2016)}]{Kosterlitz_2016}%
  \BibitemOpen
  \bibfield  {author} {\bibinfo {author} {\bibfnamefont {J.~M.}\ \bibnamefont
  {Kosterlitz}},\ }\href {\doibase 10.1088/0034-4885/79/2/026001} {\bibfield
  {journal} {\bibinfo  {journal} {Rep. Prog. Phys.}\ }\textbf {\bibinfo
  {volume} {79}},\ \bibinfo {pages} {026001} (\bibinfo {year}
  {2016})}\BibitemShut {NoStop}%
\bibitem [{\citenamefont {Hasenbusch}\ \emph
  {et~al.}(2005{\natexlab{b}})\citenamefont {Hasenbusch}, \citenamefont
  {Pelissetto},\ and\ \citenamefont {Vicari}}]{PhysRevB.72.184502}%
  \BibitemOpen
  \bibfield  {author} {\bibinfo {author} {\bibfnamefont {M.}~\bibnamefont
  {Hasenbusch}}, \bibinfo {author} {\bibfnamefont {A.}~\bibnamefont
  {Pelissetto}}, \ and\ \bibinfo {author} {\bibfnamefont {E.}~\bibnamefont
  {Vicari}},\ }\href {\doibase 10.1103/PhysRevB.72.184502} {\bibfield
  {journal} {\bibinfo  {journal} {Phys. Rev. B}\ }\textbf {\bibinfo {volume}
  {72}},\ \bibinfo {pages} {184502} (\bibinfo {year}
  {2005}{\natexlab{b}})}\BibitemShut {NoStop}%
\bibitem [{\citenamefont {Lima}\ \emph {et~al.}(2019)\citenamefont {Lima},
  \citenamefont {M{\'o}l},\ and\ \citenamefont {Costa}}]{lima2018fully}%
  \BibitemOpen
  \bibfield  {author} {\bibinfo {author} {\bibfnamefont {A.~B.}\ \bibnamefont
  {Lima}}, \bibinfo {author} {\bibfnamefont {L.~A.~S.}\ \bibnamefont
  {M{\'o}l}}, \ and\ \bibinfo {author} {\bibfnamefont {B.~V.}\ \bibnamefont
  {Costa}},\ }\href {https://doi.org/10.1007/s10955-019-02271-x} {\bibfield
  {journal} {\bibinfo  {journal} {J. Stat. Phys.}\ } (\bibinfo {year}
  {2019})}\BibitemShut {NoStop}%
\bibitem [{\citenamefont {Weber}\ and\ \citenamefont
  {Minnhagen}(1988)}]{PhysRevB.37.5986}%
  \BibitemOpen
  \bibfield  {author} {\bibinfo {author} {\bibfnamefont {H.}~\bibnamefont
  {Weber}}\ and\ \bibinfo {author} {\bibfnamefont {P.}~\bibnamefont
  {Minnhagen}},\ }\href {\doibase 10.1103/PhysRevB.37.5986} {\bibfield
  {journal} {\bibinfo  {journal} {Phys. Rev. B}\ }\textbf {\bibinfo {volume}
  {37}},\ \bibinfo {pages} {5986} (\bibinfo {year} {1988})}\BibitemShut
  {NoStop}%
\bibitem [{\citenamefont {Hasenbusch}(2005)}]{hasenbusch2005two}%
  \BibitemOpen
  \bibfield  {author} {\bibinfo {author} {\bibfnamefont {M.}~\bibnamefont
  {Hasenbusch}},\ }\href@noop {} {\bibfield  {journal} {\bibinfo  {journal} {J.
  Phys. A.}\ }\textbf {\bibinfo {volume} {38}},\ \bibinfo {pages} {5869}
  (\bibinfo {year} {2005})}\BibitemShut {NoStop}%
\bibitem [{\citenamefont {Cai}\ \emph {et~al.}(2014)\citenamefont {Cai},
  \citenamefont {Schollw\"ock},\ and\ \citenamefont
  {Pollet}}]{PhysRevLett.113.260403}%
  \BibitemOpen
  \bibfield  {author} {\bibinfo {author} {\bibfnamefont {Z.}~\bibnamefont
  {Cai}}, \bibinfo {author} {\bibfnamefont {U.}~\bibnamefont {Schollw\"ock}}, \
  and\ \bibinfo {author} {\bibfnamefont {L.}~\bibnamefont {Pollet}},\ }\href
  {\doibase 10.1103/PhysRevLett.113.260403} {\bibfield  {journal} {\bibinfo
  {journal} {Phys. Rev. Lett.}\ }\textbf {\bibinfo {volume} {113}},\ \bibinfo
  {pages} {260403} (\bibinfo {year} {2014})}\BibitemShut {NoStop}%
\bibitem [{\citenamefont {Gerster}\ \emph {et~al.}(2016)\citenamefont
  {Gerster}, \citenamefont {Rizzi}, \citenamefont {Tschirsich}, \citenamefont
  {Silvi}, \citenamefont {Fazio},\ and\ \citenamefont
  {Montangero}}]{gerster2016superfluid}%
  \BibitemOpen
  \bibfield  {author} {\bibinfo {author} {\bibfnamefont {M.}~\bibnamefont
  {Gerster}}, \bibinfo {author} {\bibfnamefont {M.}~\bibnamefont {Rizzi}},
  \bibinfo {author} {\bibfnamefont {F.}~\bibnamefont {Tschirsich}}, \bibinfo
  {author} {\bibfnamefont {P.}~\bibnamefont {Silvi}}, \bibinfo {author}
  {\bibfnamefont {R.}~\bibnamefont {Fazio}}, \ and\ \bibinfo {author}
  {\bibfnamefont {S.}~\bibnamefont {Montangero}},\ }\href@noop {} {\bibfield
  {journal} {\bibinfo  {journal} {New J. Phys.}\ }\textbf {\bibinfo {volume}
  {18}},\ \bibinfo {pages} {015015} (\bibinfo {year} {2016})}\BibitemShut
  {NoStop}%
\bibitem [{\citenamefont {Nelson}\ and\ \citenamefont
  {Kosterlitz}(1977)}]{PhysRevLett.39.1201}%
  \BibitemOpen
  \bibfield  {author} {\bibinfo {author} {\bibfnamefont {D.~R.}\ \bibnamefont
  {Nelson}}\ and\ \bibinfo {author} {\bibfnamefont {J.~M.}\ \bibnamefont
  {Kosterlitz}},\ }\href {\doibase 10.1103/PhysRevLett.39.1201} {\bibfield
  {journal} {\bibinfo  {journal} {Phys. Rev. Lett.}\ }\textbf {\bibinfo
  {volume} {39}},\ \bibinfo {pages} {1201} (\bibinfo {year}
  {1977})}\BibitemShut {NoStop}%
\bibitem [{\citenamefont {Greitemann}\ \emph {et~al.}(2015)\citenamefont
  {Greitemann}, \citenamefont {Hesselmann}, \citenamefont {Wessel},
  \citenamefont {Assaad},\ and\ \citenamefont
  {Hohenadler}}]{PhysRevB.92.245132}%
  \BibitemOpen
  \bibfield  {author} {\bibinfo {author} {\bibfnamefont {J.}~\bibnamefont
  {Greitemann}}, \bibinfo {author} {\bibfnamefont {S.}~\bibnamefont
  {Hesselmann}}, \bibinfo {author} {\bibfnamefont {S.}~\bibnamefont {Wessel}},
  \bibinfo {author} {\bibfnamefont {F.~F.}\ \bibnamefont {Assaad}}, \ and\
  \bibinfo {author} {\bibfnamefont {M.}~\bibnamefont {Hohenadler}},\ }\href
  {\doibase 10.1103/PhysRevB.92.245132} {\bibfield  {journal} {\bibinfo
  {journal} {Phys. Rev. B}\ }\textbf {\bibinfo {volume} {92}},\ \bibinfo
  {pages} {245132} (\bibinfo {year} {2015})}\BibitemShut {NoStop}%
\bibitem [{\citenamefont {Voit}\ and\ \citenamefont {Schulz}(1986)}]{Voit86}%
  \BibitemOpen
  \bibfield  {author} {\bibinfo {author} {\bibfnamefont {J.}~\bibnamefont
  {Voit}}\ and\ \bibinfo {author} {\bibfnamefont {H.~J.}\ \bibnamefont
  {Schulz}},\ }\href {\doibase 10.1103/PhysRevB.34.7429} {\bibfield  {journal}
  {\bibinfo  {journal} {Phys. Rev. B}\ }\textbf {\bibinfo {volume} {34}},\
  \bibinfo {pages} {7429} (\bibinfo {year} {1986})}\BibitemShut {NoStop}%
\bibitem [{\citenamefont {Kuboki}\ and\ \citenamefont
  {Fukuyama}(1987)}]{kuboki1987spin}%
  \BibitemOpen
  \bibfield  {author} {\bibinfo {author} {\bibfnamefont {K.}~\bibnamefont
  {Kuboki}}\ and\ \bibinfo {author} {\bibfnamefont {H.}~\bibnamefont
  {Fukuyama}},\ }\href@noop {} {\bibfield  {journal} {\bibinfo  {journal} {J.
  Phys. Soc. Jpn.}\ }\textbf {\bibinfo {volume} {56}},\ \bibinfo {pages} {3126}
  (\bibinfo {year} {1987})}\BibitemShut {NoStop}%
\bibitem [{\citenamefont {Raas}\ \emph {et~al.}(2002)\citenamefont {Raas},
  \citenamefont {L\"ow}, \citenamefont {Uhrig},\ and\ \citenamefont
  {K\"uhne}}]{PhysRevB.65.144438}%
  \BibitemOpen
  \bibfield  {author} {\bibinfo {author} {\bibfnamefont {C.}~\bibnamefont
  {Raas}}, \bibinfo {author} {\bibfnamefont {U.}~\bibnamefont {L\"ow}},
  \bibinfo {author} {\bibfnamefont {G.~S.}\ \bibnamefont {Uhrig}}, \ and\
  \bibinfo {author} {\bibfnamefont {R.~W.}\ \bibnamefont {K\"uhne}},\ }\href
  {\doibase 10.1103/PhysRevB.65.144438} {\bibfield  {journal} {\bibinfo
  {journal} {Phys. Rev. B}\ }\textbf {\bibinfo {volume} {65}},\ \bibinfo
  {pages} {144438} (\bibinfo {year} {2002})}\BibitemShut {NoStop}%
\bibitem [{\citenamefont {Wei\ss{}e}\ \emph {et~al.}(2006)\citenamefont
  {Wei\ss{}e}, \citenamefont {Hager}, \citenamefont {Bishop},\ and\
  \citenamefont {Fehske}}]{PhysRevB.74.214426}%
  \BibitemOpen
  \bibfield  {author} {\bibinfo {author} {\bibfnamefont {A.}~\bibnamefont
  {Wei\ss{}e}}, \bibinfo {author} {\bibfnamefont {G.}~\bibnamefont {Hager}},
  \bibinfo {author} {\bibfnamefont {A.~R.}\ \bibnamefont {Bishop}}, \ and\
  \bibinfo {author} {\bibfnamefont {H.}~\bibnamefont {Fehske}},\ }\href
  {\doibase 10.1103/PhysRevB.74.214426} {\bibfield  {journal} {\bibinfo
  {journal} {Phys. Rev. B}\ }\textbf {\bibinfo {volume} {74}},\ \bibinfo
  {pages} {214426} (\bibinfo {year} {2006})}\BibitemShut {NoStop}%
\bibitem [{\citenamefont {Aoki}(1984)}]{PhysRevD.30.2653}%
  \BibitemOpen
  \bibfield  {author} {\bibinfo {author} {\bibfnamefont {S.}~\bibnamefont
  {Aoki}},\ }\href {\doibase 10.1103/PhysRevD.30.2653} {\bibfield  {journal}
  {\bibinfo  {journal} {Phys. Rev. D}\ }\textbf {\bibinfo {volume} {30}},\
  \bibinfo {pages} {2653} (\bibinfo {year} {1984})}\BibitemShut {NoStop}%
\bibitem [{\citenamefont {{Affleck}}\ \emph {et~al.}(1989)\citenamefont
  {{Affleck}}, \citenamefont {{Gepner}}, \citenamefont {{Schulz}},\ and\
  \citenamefont {{Ziman}}}]{1989JPhA...22..511A}%
  \BibitemOpen
  \bibfield  {author} {\bibinfo {author} {\bibfnamefont {I.}~\bibnamefont
  {{Affleck}}}, \bibinfo {author} {\bibfnamefont {D.}~\bibnamefont {{Gepner}}},
  \bibinfo {author} {\bibfnamefont {H.~J.}\ \bibnamefont {{Schulz}}}, \ and\
  \bibinfo {author} {\bibfnamefont {T.}~\bibnamefont {{Ziman}}},\ }\href
  {\doibase 10.1088/0305-4470/22/5/015} {\bibfield  {journal} {\bibinfo
  {journal} {Journal of Physics A: Mathematical General}\ }\textbf {\bibinfo
  {volume} {22}},\ \bibinfo {pages} {511} (\bibinfo {year} {1989})}\BibitemShut
  {NoStop}%
\bibitem [{\citenamefont {Gross}\ and\ \citenamefont
  {Neveu}(1974)}]{PhysRevD.10.3235}%
  \BibitemOpen
  \bibfield  {author} {\bibinfo {author} {\bibfnamefont {D.~J.}\ \bibnamefont
  {Gross}}\ and\ \bibinfo {author} {\bibfnamefont {A.}~\bibnamefont {Neveu}},\
  }\href {\doibase 10.1103/PhysRevD.10.3235} {\bibfield  {journal} {\bibinfo
  {journal} {Phys. Rev. D}\ }\textbf {\bibinfo {volume} {10}},\ \bibinfo
  {pages} {3235} (\bibinfo {year} {1974})}\BibitemShut {NoStop}%
\bibitem [{\citenamefont {Mudry}(2014)}]{mudry2014lecture}%
  \BibitemOpen
  \bibfield  {author} {\bibinfo {author} {\bibfnamefont {C.}~\bibnamefont
  {Mudry}},\ }\href@noop {} {\emph {\bibinfo {title} {Lecture notes on field
  theory in condensed matter physics}}}\ (\bibinfo  {publisher} {World
  Scientific Publishing Company},\ \bibinfo {year} {2014})\BibitemShut
  {NoStop}%
\bibitem [{\citenamefont {Goldstone}\ and\ \citenamefont
  {Wilczek}(1981)}]{PhysRevLett.47.986}%
  \BibitemOpen
  \bibfield  {author} {\bibinfo {author} {\bibfnamefont {J.}~\bibnamefont
  {Goldstone}}\ and\ \bibinfo {author} {\bibfnamefont {F.}~\bibnamefont
  {Wilczek}},\ }\href {\doibase 10.1103/PhysRevLett.47.986} {\bibfield
  {journal} {\bibinfo  {journal} {Phys. Rev. Lett.}\ }\textbf {\bibinfo
  {volume} {47}},\ \bibinfo {pages} {986} (\bibinfo {year} {1981})}\BibitemShut
  {NoStop}%
\bibitem [{\citenamefont {Ryu}\ \emph {et~al.}(2009)\citenamefont {Ryu},
  \citenamefont {Mudry}, \citenamefont {Hou},\ and\ \citenamefont
  {Chamon}}]{PhysRevB.80.205319}%
  \BibitemOpen
  \bibfield  {author} {\bibinfo {author} {\bibfnamefont {S.}~\bibnamefont
  {Ryu}}, \bibinfo {author} {\bibfnamefont {C.}~\bibnamefont {Mudry}}, \bibinfo
  {author} {\bibfnamefont {C.-Y.}\ \bibnamefont {Hou}}, \ and\ \bibinfo
  {author} {\bibfnamefont {C.}~\bibnamefont {Chamon}},\ }\href {\doibase
  10.1103/PhysRevB.80.205319} {\bibfield  {journal} {\bibinfo  {journal} {Phys.
  Rev. B}\ }\textbf {\bibinfo {volume} {80}},\ \bibinfo {pages} {205319}
  (\bibinfo {year} {2009})}\BibitemShut {NoStop}%
\bibitem [{\citenamefont {Sticlet}\ \emph {et~al.}(2014)\citenamefont
  {Sticlet}, \citenamefont {Seabra}, \citenamefont {Pollmann},\ and\
  \citenamefont {Cayssol}}]{PhysRevB.89.115430}%
  \BibitemOpen
  \bibfield  {author} {\bibinfo {author} {\bibfnamefont {D.}~\bibnamefont
  {Sticlet}}, \bibinfo {author} {\bibfnamefont {L.}~\bibnamefont {Seabra}},
  \bibinfo {author} {\bibfnamefont {F.}~\bibnamefont {Pollmann}}, \ and\
  \bibinfo {author} {\bibfnamefont {J.}~\bibnamefont {Cayssol}},\ }\href
  {\doibase 10.1103/PhysRevB.89.115430} {\bibfield  {journal} {\bibinfo
  {journal} {Phys. Rev. B}\ }\textbf {\bibinfo {volume} {89}},\ \bibinfo
  {pages} {115430} (\bibinfo {year} {2014})}\BibitemShut {NoStop}%
\bibitem [{\citenamefont {Kuno}(2019)}]{PhysRevB.99.064105}%
  \BibitemOpen
  \bibfield  {author} {\bibinfo {author} {\bibfnamefont {Y.}~\bibnamefont
  {Kuno}},\ }\href {\doibase 10.1103/PhysRevB.99.064105} {\bibfield  {journal}
  {\bibinfo  {journal} {Phys. Rev. B}\ }\textbf {\bibinfo {volume} {99}},\
  \bibinfo {pages} {064105} (\bibinfo {year} {2019})}\BibitemShut {NoStop}%
\bibitem [{\citenamefont {Beyl}\ \emph {et~al.}()\citenamefont {Beyl},
  \citenamefont {Hohenadler}, \citenamefont {Goth},\ and\ \citenamefont
  {Assaad}}]{Be.Ho.Go.As.2019}%
  \BibitemOpen
  \bibfield  {author} {\bibinfo {author} {\bibfnamefont {S.}~\bibnamefont
  {Beyl}}, \bibinfo {author} {\bibfnamefont {M.}~\bibnamefont {Hohenadler}},
  \bibinfo {author} {\bibfnamefont {F.}~\bibnamefont {Goth}}, \ and\ \bibinfo
  {author} {\bibfnamefont {F.~F.}\ \bibnamefont {Assaad}},\ }\href@noop {}
  {\bibinfo  {journal} {in preparation}\ }\BibitemShut {NoStop}%
\bibitem [{\citenamefont {Sandvik}\ \emph {et~al.}(1997)\citenamefont
  {Sandvik}, \citenamefont {Singh},\ and\ \citenamefont
  {Campbell}}]{SandvikPhononsa}%
  \BibitemOpen
\bibfield  {journal} {  }\bibfield  {author} {\bibinfo {author} {\bibfnamefont
  {A.~W.}\ \bibnamefont {Sandvik}}, \bibinfo {author} {\bibfnamefont
  {R.~R.~P.}\ \bibnamefont {Singh}}, \ and\ \bibinfo {author} {\bibfnamefont
  {D.~K.}\ \bibnamefont {Campbell}},\ }\href@noop {} {\bibfield  {journal}
  {\bibinfo  {journal} {Phys. Rev. B}\ }\textbf {\bibinfo {volume} {56}},\
  \bibinfo {pages} {14510} (\bibinfo {year} {1997})}\BibitemShut {NoStop}%
\bibitem [{\citenamefont {Weber}\ \emph {et~al.}(2016)\citenamefont {Weber},
  \citenamefont {Assaad},\ and\ \citenamefont
  {Hohenadler}}]{PhysRevB.94.245138}%
  \BibitemOpen
  \bibfield  {author} {\bibinfo {author} {\bibfnamefont {M.}~\bibnamefont
  {Weber}}, \bibinfo {author} {\bibfnamefont {F.~F.}\ \bibnamefont {Assaad}}, \
  and\ \bibinfo {author} {\bibfnamefont {M.}~\bibnamefont {Hohenadler}},\
  }\href {\doibase 10.1103/PhysRevB.94.245138} {\bibfield  {journal} {\bibinfo
  {journal} {Phys. Rev. B}\ }\textbf {\bibinfo {volume} {94}},\ \bibinfo
  {pages} {245138} (\bibinfo {year} {2016})}\BibitemShut {NoStop}%
\bibitem [{\citenamefont {Weber}\ \emph {et~al.}(2018)\citenamefont {Weber},
  \citenamefont {Assaad},\ and\ \citenamefont
  {Hohenadler}}]{PhysRevB.98.235117}%
  \BibitemOpen
  \bibfield  {author} {\bibinfo {author} {\bibfnamefont {M.}~\bibnamefont
  {Weber}}, \bibinfo {author} {\bibfnamefont {F.~F.}\ \bibnamefont {Assaad}}, \
  and\ \bibinfo {author} {\bibfnamefont {M.}~\bibnamefont {Hohenadler}},\
  }\href {\doibase 10.1103/PhysRevB.98.235117} {\bibfield  {journal} {\bibinfo
  {journal} {Phys. Rev. B}\ }\textbf {\bibinfo {volume} {98}},\ \bibinfo
  {pages} {235117} (\bibinfo {year} {2018})}\BibitemShut {NoStop}%
\bibitem [{\citenamefont {Het{\'e}nyi}(2014)}]{hetenyi2014drude}%
  \BibitemOpen
  \bibfield  {author} {\bibinfo {author} {\bibfnamefont {B.}~\bibnamefont
  {Het{\'e}nyi}},\ }\href@noop {} {\bibfield  {journal} {\bibinfo  {journal}
  {J. Phys. Soc. Jpn.}\ }\textbf {\bibinfo {volume} {83}},\ \bibinfo {pages}
  {034711} (\bibinfo {year} {2014})}\BibitemShut {NoStop}%
\bibitem [{\citenamefont {Pollock}\ and\ \citenamefont
  {Ceperley}(1987)}]{PhysRevB.36.8343}%
  \BibitemOpen
  \bibfield  {author} {\bibinfo {author} {\bibfnamefont {E.~L.}\ \bibnamefont
  {Pollock}}\ and\ \bibinfo {author} {\bibfnamefont {D.~M.}\ \bibnamefont
  {Ceperley}},\ }\href {\doibase 10.1103/PhysRevB.36.8343} {\bibfield
  {journal} {\bibinfo  {journal} {Phys. Rev. B}\ }\textbf {\bibinfo {volume}
  {36}},\ \bibinfo {pages} {8343} (\bibinfo {year} {1987})}\BibitemShut
  {NoStop}%
\bibitem [{\citenamefont {Sandvik}\ \emph {et~al.}(2004)\citenamefont
  {Sandvik}, \citenamefont {Balents},\ and\ \citenamefont
  {Campbell}}]{Sa.Ba.Ca.04}%
  \BibitemOpen
  \bibfield  {author} {\bibinfo {author} {\bibfnamefont {A.~W.}\ \bibnamefont
  {Sandvik}}, \bibinfo {author} {\bibfnamefont {L.}~\bibnamefont {Balents}}, \
  and\ \bibinfo {author} {\bibfnamefont {D.~K.}\ \bibnamefont {Campbell}},\
  }\href@noop {} {\bibfield  {journal} {\bibinfo  {journal} {Phys. Rev. Lett.}\
  }\textbf {\bibinfo {volume} {92}},\ \bibinfo {pages} {236401} (\bibinfo
  {year} {2004})}\BibitemShut {NoStop}%
\bibitem [{\citenamefont {Pelissetto}\ and\ \citenamefont
  {Vicari}(2013)}]{PhysRevE.87.032105}%
  \BibitemOpen
  \bibfield  {author} {\bibinfo {author} {\bibfnamefont {A.}~\bibnamefont
  {Pelissetto}}\ and\ \bibinfo {author} {\bibfnamefont {E.}~\bibnamefont
  {Vicari}},\ }\href {\doibase 10.1103/PhysRevE.87.032105} {\bibfield
  {journal} {\bibinfo  {journal} {Phys. Rev. E}\ }\textbf {\bibinfo {volume}
  {87}},\ \bibinfo {pages} {032105} (\bibinfo {year} {2013})}\BibitemShut
  {NoStop}%
\bibitem [{\citenamefont {Laflorencie}\ \emph {et~al.}(2001)\citenamefont
  {Laflorencie}, \citenamefont {Capponi},\ and\ \citenamefont
  {S{\o}rensen}}]{laflorencie2001finite}%
  \BibitemOpen
  \bibfield  {author} {\bibinfo {author} {\bibfnamefont {N.}~\bibnamefont
  {Laflorencie}}, \bibinfo {author} {\bibfnamefont {S.}~\bibnamefont
  {Capponi}}, \ and\ \bibinfo {author} {\bibfnamefont {E.~S.}\ \bibnamefont
  {S{\o}rensen}},\ }\href@noop {} {\bibfield  {journal} {\bibinfo  {journal}
  {Eur. Phys. J. B}\ }\textbf {\bibinfo {volume} {24}},\ \bibinfo {pages} {77}
  (\bibinfo {year} {2001})}\BibitemShut {NoStop}%
\bibitem [{\citenamefont {Melko}\ \emph {et~al.}(2004)\citenamefont {Melko},
  \citenamefont {Sandvik},\ and\ \citenamefont
  {Scalapino}}]{PhysRevB.69.014509}%
  \BibitemOpen
  \bibfield  {author} {\bibinfo {author} {\bibfnamefont {R.~G.}\ \bibnamefont
  {Melko}}, \bibinfo {author} {\bibfnamefont {A.~W.}\ \bibnamefont {Sandvik}},
  \ and\ \bibinfo {author} {\bibfnamefont {D.~J.}\ \bibnamefont {Scalapino}},\
  }\href {\doibase 10.1103/PhysRevB.69.014509} {\bibfield  {journal} {\bibinfo
  {journal} {Phys. Rev. B}\ }\textbf {\bibinfo {volume} {69}},\ \bibinfo
  {pages} {014509} (\bibinfo {year} {2004})}\BibitemShut {NoStop}%
\bibitem [{\citenamefont {Benfatto}\ \emph {et~al.}(2013)\citenamefont
  {Benfatto}, \citenamefont {Castellani},\ and\ \citenamefont
  {Giamarchi}}]{BKTsinegordon}%
  \BibitemOpen
  \bibfield  {author} {\bibinfo {author} {\bibfnamefont {L.}~\bibnamefont
  {Benfatto}}, \bibinfo {author} {\bibfnamefont {C.}~\bibnamefont
  {Castellani}}, \ and\ \bibinfo {author} {\bibfnamefont {T.}~\bibnamefont
  {Giamarchi}},\ }in\ \href@noop {} {\emph {\bibinfo {booktitle} {40 years of
  {B}erezinskii-{K}osterlitz-{T}houless theory}}},\ \bibinfo {editor} {edited
  by\ \bibinfo {editor} {\bibfnamefont {J.~V.}\ \bibnamefont {Jos}}}\ (\bibinfo
   {publisher} {World Scientific},\ \bibinfo {year} {2013})\BibitemShut
  {NoStop}%
\end{thebibliography}


\end{document}